\documentclass[12pt,preprint]{aastex}

\def\F0{F_{\rm 0}}
\def\t0{t_{\rm 0}}

\newcommand{\ltsima} {$\; \buildrel < \over \sim \;$}
\newcommand{\gtsima} {$\; \buildrel > \over \sim \;$}
\newcommand{\lta} {\lower.5ex\hbox{\ltsima}}
\newcommand{\gta} {\lower.5ex\hbox{\gtsima}}

\slugcomment{Date: \today}

\begin{document}

\title{Gamma-Ray Burst Spectral Correlations: Photospheric and
Injection Effects}

\author{Felix Ryde}
\affil{Stockholm Observatory, AlbaNova, SE-106 91 Stockholm,
Sweden and Department of Astronomy \& Astrophysics, Pennsylvania
State University, University Park, PA 16802, USA}

\author{Claes-Ingvar Bj\"ornsson}
\affil{Stockholm Observatory, AlbaNova, SE-106 91 Stockholm,
 Sweden}

\author{Yuki Kaneko}
\affil{Universities Space Research Association, National Space
Science and Technology Center, Huntsville, AL, 35805, USA}

\author{Peter M\'esz\'aros}
\affil{Department of Astronomy \& Astrophysics, Pennsylvania State
University, University Park, PA 16802, USA and Department of
Physics, Pennsylvania State University, University Park, PA 16802,
USA}

\author{Robert Preece}
\affil{Physics Department, University of Alabama in Huntsville,
National Space Science and Technology Center, Huntsville, AL
35899, USA}

\author{Milan Battelino}
\affil{Stockholm Observatory, AlbaNova, SE-106 91 Stockholm,
 Sweden}

\begin{abstract}
We present a physical framework that can account for most of the
observed spectral properties of the prompt gamma-ray burst
emission. This includes the variety of spectral shapes, shape
evolutions, and spectral correlations between flux and spectral
peak,  within bursts described by Borgonovo \& Ryde, and among
bursts described by Amati/Ghirlanda. In our proposed model the
spectral peak is given by the photospheric emission from a
relativistic outflow for which the horizon length is much smaller
than the radial width. The observed duration of the thermal flash
will be given by the radial light-crossing time. This then gives
that the typical emission site is at $\sim 10^{11}$cm from the
center, with a Lorentz factor of $\sim 300$. This emission is
accompanied by non-thermal emission from dissipation locations
outside the photosphere. The relative strength of these two
components depend on injection effects at the central engine
leading to varying relative location of the saturation and
photospheric radii. The total emission can then reproduce the
observed variety. The spectral correlations are found by assuming
that the amount of energy dissipated depends non-linearly on the
averaged particle density. Beside the spectral correlations this
also gives a description of how the relative strength of the
thermal component varies with temperature within a burst.
\end{abstract}


\section{Introduction}

The emission below the spectral peak in gamma-ray bursts (GRBs),
which typically lies at around $200 - 300$ keV, is indicative of
the radiative process in charge of the spectrum formation. In this
sense a truly crucial observation is that the power-law slope of
the sub-peak emission in a fair fraction of observed,
instantaneous spectra are harder than what is expected from
optically-thin synchrotron emission \citep{preece00}. Indeed, the
{\it Compton Gamma Ray Observatory} BATSE instrument [25--2000
keV] recorded spectra which were as hard as a thermal emission
throughout the burst \citep{ryde04} as well as during the initial
interval of bursts \citep{ghirlanda, ryde04, kaneko05, bosnjak}.
Furthermore, {\it Beppo-SAX} [2-700 keV] detected several bursts
which exhibited very hard low-energy spectra (see e.g.
\citet{frontera00}).

Moreover, observations made by {\it Ginga} [$\sim$ 2-400 keV]
\citep{strohmayer} indicated a substantial number of bursts with
breaks below 10 keV, i.e., below the observable range of BATSE.
The authors suggested that bursts might have two different
spectral breaks, one in the BATSE range and one below this, close
to 5 keV. Similar results are reported by {\it Swift} [0.3-500
keV, with {\it KONUS}] for GRB050820A (Osborne et al. 2005) whose
spectrum indicates that there can be an additional spectral break
around 10 keV. These evidence indicate that the high-energy
spectra of GRBs are more complicated than that of a single broken
power-law. Similarly, the need for additional high-energy spectral
components beyond 1 MeV in several bursts (e.g. \citet{hurley,
atkins, gonz}) reinforces the same conclusions.

Another crucial point was raised by Ryde (2005) who argued that a
strong thermal component could very well be a common feature in
the $\gamma$-ray spectra of GRBs. Most of the strong, pulsed
bursts analyzed could be described by a two-component model
consisting of a thermal component, modelled by a Planck function,
and a non-thermal component, modelled by a power-law function, the
sum of which creates the observed spectral shapes and evolutions.
The typical peak in a GRB spectrum is thus argued to be caused by
a thermal component. The behaviors of the two separate components
on their own are, in general, similar from burst to burst, even
though the combined behavior leads to the large variety in
spectral evolutions that is observed. The typical behavior of the
temperature over a single pulse episode is, with a few exceptions,
that it decreases monotonically as a broken power-law in time. The
non-thermal emission component, which is parameterized by a
power-law photon-index $s \equiv d\,\log N_{\rm E}/d\,\log E$,
where $N_E$ is the photon flux, exhibits a typical decrease in $s$
from approximately $-1.5$ to approximately $-2$. Other behaviors
do occasionally exist (see Ryde et al. in prep.) These results
also hold for bursts which are consistent with optically-thin
synchrotron emission, in the sense that spectral fits with a
broken power-law function gives a photon-index of the low-energy
power-law which is smaller (softer) than $\alpha = -2/3$.

Such thermal--non-thermal hybrid emission is indeed common in
astrophysical sources and is a natural consequence of heating of a
background plasma and acceleration of particles. This is, for
instance, typical for solar flares, accretion disks and cluster of
galaxies (see, e.g., Petrosian \& Liu 2004). In the case of GRBs,
the thermal component is thought to stem from the fireball
photosphere while the non-thermal component stems from accelerated
electrons beyond the photospheric radius, in the optically-thin
region.  We will denote this model as the photosphere model since
the main feature of it, its energy flux peak, and spectral
evolution is given by the thermal component. This certainly
emanates from the photosphere, where the optical depth is unity,
even though the finally observed emission can be altered due to
scattering and diffusion effects.

If we are indeed observing the photospheric, thermal emission, as
suggested by the analyses above, it simplifies the physical
interpretation since we are detecting the fireball and its
evolution directly. This is in contrast to the internal shocks
which give us indirect information through the dissipation
processes, randomly distributed through out the fireball wind,
which by necessity complicates the interpretation. However, the
origin of the thermal component and its observed evolution is not
obvious. Strong photospheric components were predicted in the
spectra of GRBs, in kinetic models by \citet{MR00, MRR, DM,
ryde04}, and in Poynting-flux dominated models by  e.g.
\citet{DS}. More recently, discussions on various scenarios and
theoretical aspects have been given by, e.g., \citet{RM05,
thomson, asaf, TMR06}. In this paper we will assume that the
thermal component is predominantly from close to the photosphere
and that the evolution we detect is mainly due to change in the
property of the photosphere. First, we discuss the observational
and statistical significance of the thermal fits in \S
\ref{sec:sign}. We then discuss some basic properties of the
fireball evolution that we can learn from the observed
$\gamma$-ray behavior in \S \ref{sec:theory} and derive several of
the correlations that have been observed for bursts. We discuss
our results in \S \ref{sec:discussion} and comment on alternative
models.

\section{Observational Motivation for the Photosphere Model}
\label{sec:sign}

The photosphere model, which combines a Planck spectrum (see
Appendix A) with with a power-law, $N_{\rm E} \propto E^{s}$, is
an alternative way of fitting the observed GRB spectra as compared
to the commonly used Band et al. (1993) function. The latter
function consists of two power-laws, exponentially connected at
the break energy. These two models therefore have the same number
of parameters; temperature, power-law slope and two normalizations
compared to two power-law slopes, a peak energy and a
normalization. The two alternative fitting functions often gives
similarly good fits to the data. However, the use of the
photosphere model will lead to several important changes in the
interpretation of the spectra and their evolution. First we
discuss the significance of the fits and later discuss the change
in interpretation.

\subsection{Significance of the hard spectral slopes}

\begin{figure}[!ht]
\includegraphics[width=1.0\textwidth]{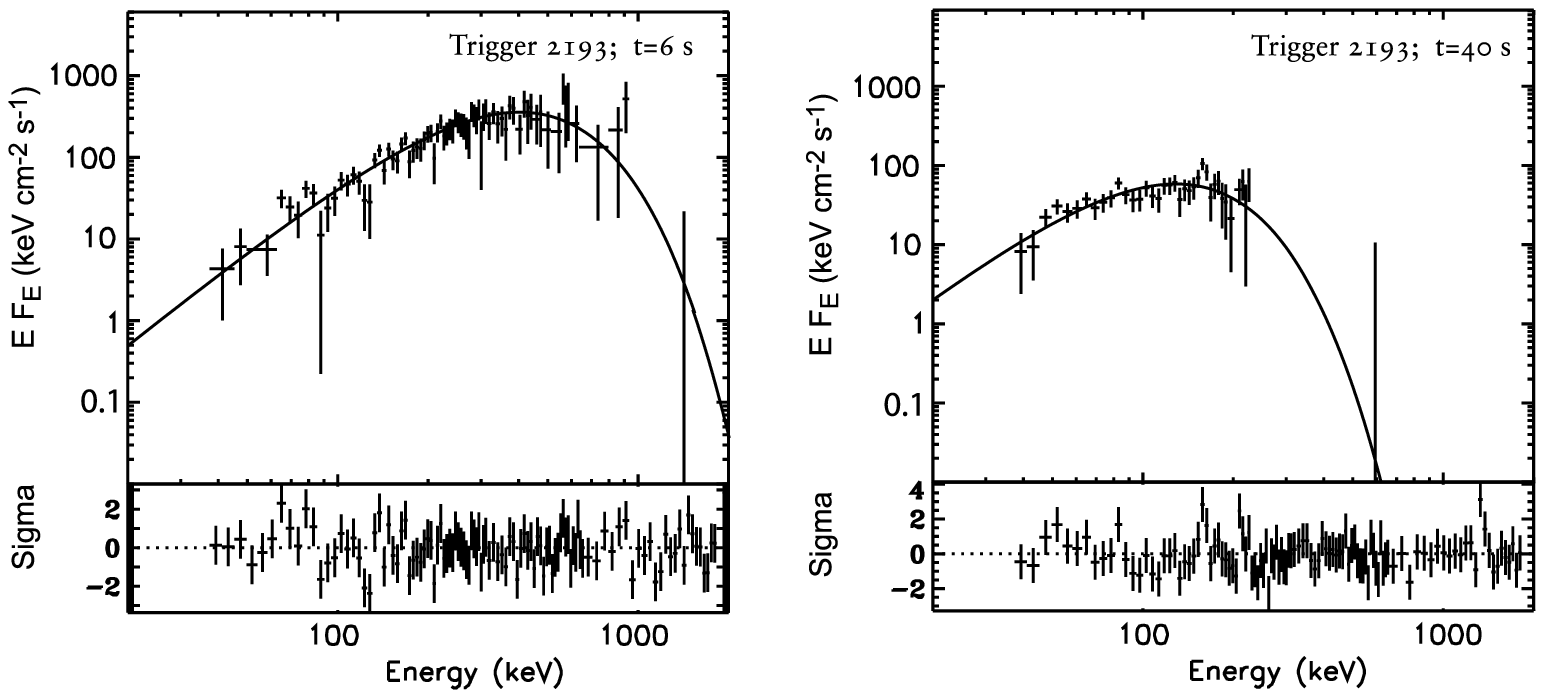}
\caption{Spectra with hard sub-peak slopes.  Two time-resolved
spectra from GRB930214 (BATSE trigger 2193) from 6 and 40 seconds
after the trigger (sub-second integration time). Note that the
spectra are fitted well with a Planck function, both in the
Rayleigh-Jeans portion of the spectrum with $\alpha=+1$ and in the
Wien portion with a fast voidance of flux. The temperature has
changed between the measurements (see also \citet{ryde04}). The
spectral data points have been rebinned to a higher
signal-to-noise ratio to increase clarity. The original energy
resolution is kept in the fitting and in the residual plots. }
 \label{fig:2193}
\end{figure}

Since synchrotron emission is a very efficient radiation mechanism
and is successful in describing the afterglows of GRBs (e.g.
\citet{compactness}), it has often been assumed that it
exclusively can describe the prompt phase as well \citep{katz,
Tavani}. The main observational question is therefore whether the
hard, sub-peak slopes that have been detected are statistically
incompatible with such an interpretation. An optically-thin
synchrotron model with the electrons having an isotropic
distribution of pitch angles yields a sub-peak power-law with
index $\alpha < -2/3$ in the slow cooling regime and $-3/2$ in the
fast cooling regime. To answer this question, we revisit the burst
of GRB930214 (BATSE trigger \# 2193) which is thermal through out
its duration \citep{ryde04}. Time-resolved spectra from two
different times are shown in Figure \ref{fig:2193}, where the
spectra are plotted in $E F_{\rm E} \equiv E^2 N_{\rm E}$ with
$F_{\rm E}$ being the energy flux. The best fit is found with a
Planck function. The averaged, reduced $\chi^2$-value for all the
40 time-resolved spectra fitted with the Planck function, that
varies in temperature and amplitude, is $< \chi^2_\nu > = 0.998$
for 4687 degrees of freedom (dof) leading to a statistical
probability value ($p$-value) of $0.53$. The spectrum in the
left-hand panel in Figure \ref{fig:2193} is shown as a typical
example and has $\chi^2_\nu ( {\rm dof})  = 0.857 (109)$ with
$p=0.86$. Fitting this spectrum with a synchrotron spectrum with a
sub-peak slope of $\alpha = -2/3$ (one additional parameter)
yields $\chi^2_\nu ({\rm dof}) = 1.29 (108)$ with $p=0.02$. These
$p$-values show unequivocally that the Planck function gives a
significantly better fit, this even in spite of the fact that it
has one parameter less.

Because the optically-thin synchrotron value of $\alpha = -2/3$ is
the asymptotic value of the power-law slope, the finite energy
range of the observations could have an effect. Another way of
quantifying the hardness of the sub-peak emission is therefore to
measure the tangential slope at the lowest possible energy to
place a lower limit on the asymptotes. Since the $\alpha$ index of
the \citet{band93} function is determined at $E = 0$ (the
asymptote), the actual low-energy power-law indices represented by
data with a finite energy range will always be softer (smaller)
than the $\alpha$ determined from the fit. Therefore, the actual
low-energy spectral behavior of the data can be more accurately
described by an ``effective" $\alpha$ ($\alpha_{\rm eff}$)
\citep{lod, kan06}. The $\alpha_{\rm eff}$ is the logarithmic
tangential slope of the fitted Band function at 25 keV, which is
the lower energy bound of BATSE Large Area Detectors (LADs).
Because $\alpha_{\rm eff} < \alpha$ for most $\alpha$ values, if
$\alpha_{\rm eff}$ significantly exceeds the synchrotron value,
the asymptotic low-energy power-law (outside the data energy
range) will most likely have an index that also violates the
synchrotron limit. Moreover, as an alternative to the
\citet{band93} function a smoothly-broken power-law (SBPL) model
has been successfully used to fit GRB spectra \citep{wingspan,
ryde99}. This model consists of two power laws joined at a certain
energy, and a break scale (i.e. a curvature) is determined at the
break energy. Thus, unlike the $\alpha$ of the Band function, the
low-energy power-law index ($\lambda_1$) of this model
characterizes the power law represented by the data, unless the
break scale is unreasonably large and/or the break energy is very
close to the lower energy bound. When LAD data are analyzed with
both models, the $\alpha_{\rm eff}$ of the Band function and the
$\lambda_1$ of the SBPL model are found to be consistent, if both
fits are reasonably good \citep{kan06}. Therefore, by using either
the $\alpha_{\rm eff}$ or $\lambda_1$ values, lower limits on the
asymptotic power-laws can be imposed. Taking the above-mentioned
burst, GRB930214 (\# 2193), again as an example, all time-resolved
spectra within the burst resulted in either $\alpha_{\rm eff}$ or
$\lambda_1 > -2/3$ by at least 4.5$\sigma$ with acceptable fits
(resulting $\chi^2$ within 2$\sigma$ in the $\chi^2$ probability
distribution for the given degrees of freedom). The low-energy
spectral behavior is thus significantly inconsistent with the
synchrotron spectrum, for the entire duration of the burst.

\begin{figure}[!ht]
\includegraphics[width=1.0\textwidth]{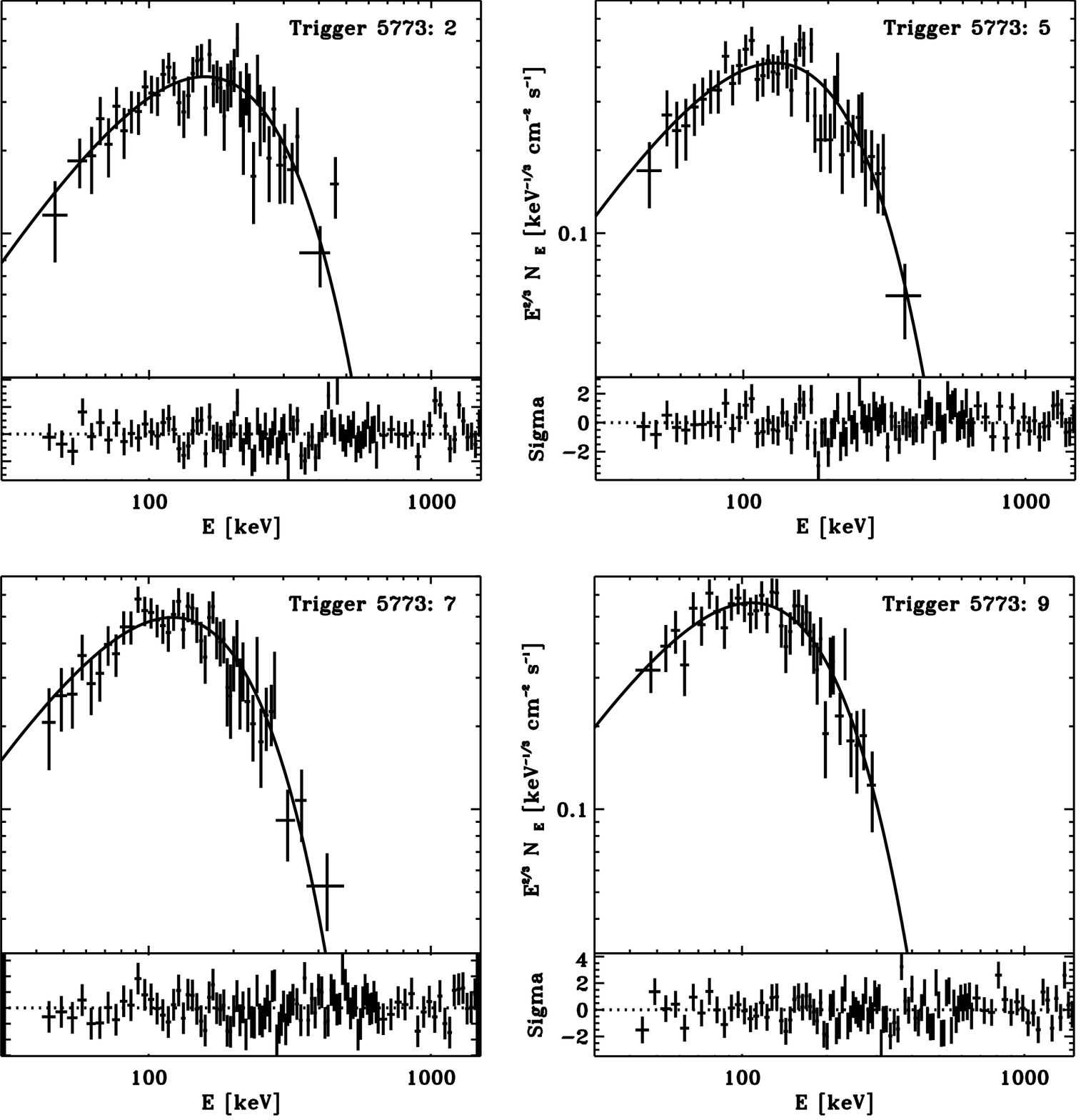}
\caption{Four time-resolved spectra of GRB 970111 (\#5773) during
the first 5 seconds of the burst. The numbers in the upper
right-hand corner refer to the sequel number of time-bin used (see
figure 13 in Ryde 2004). Note that the ordinate is  given by
$E^{2/3} N_{\rm E}$, which for instance  has the consequence that
optically-thin synchrotron emission ($\alpha = -2/3$) will appear
as a horizontal line in the plot. The low-energy slope of these
spectra are clearly significantly harder than this, which is also
seen from the analysis in the text.}
 \label{fig:5773}
\end{figure}

More commonly, bursts only exhibit short intervals of their
duration during which the spectra are thermal, most often during
an initial interval. An example of such a period is the first 5
seconds of GRB970111 (\# 5773).  Four of these spectra are shown
in figure \ref{fig:5773}. Here the spectra are plotted as $E^{2/3}
N_{\rm E}$. In such a plot the optically-thin synchrotron slope is
a horizontal line. It is clearly seen from the figure that these
spectra violate such an interpretation. As mentioned in the
introduction these conclusions are reinforced by the observation
of similarly hard spectral slopes in this burst by instruments on
{\it Beppo-SAX} \citep{frontera00}.

\begin{figure}[!ht]
\includegraphics[width=1.0\textwidth]{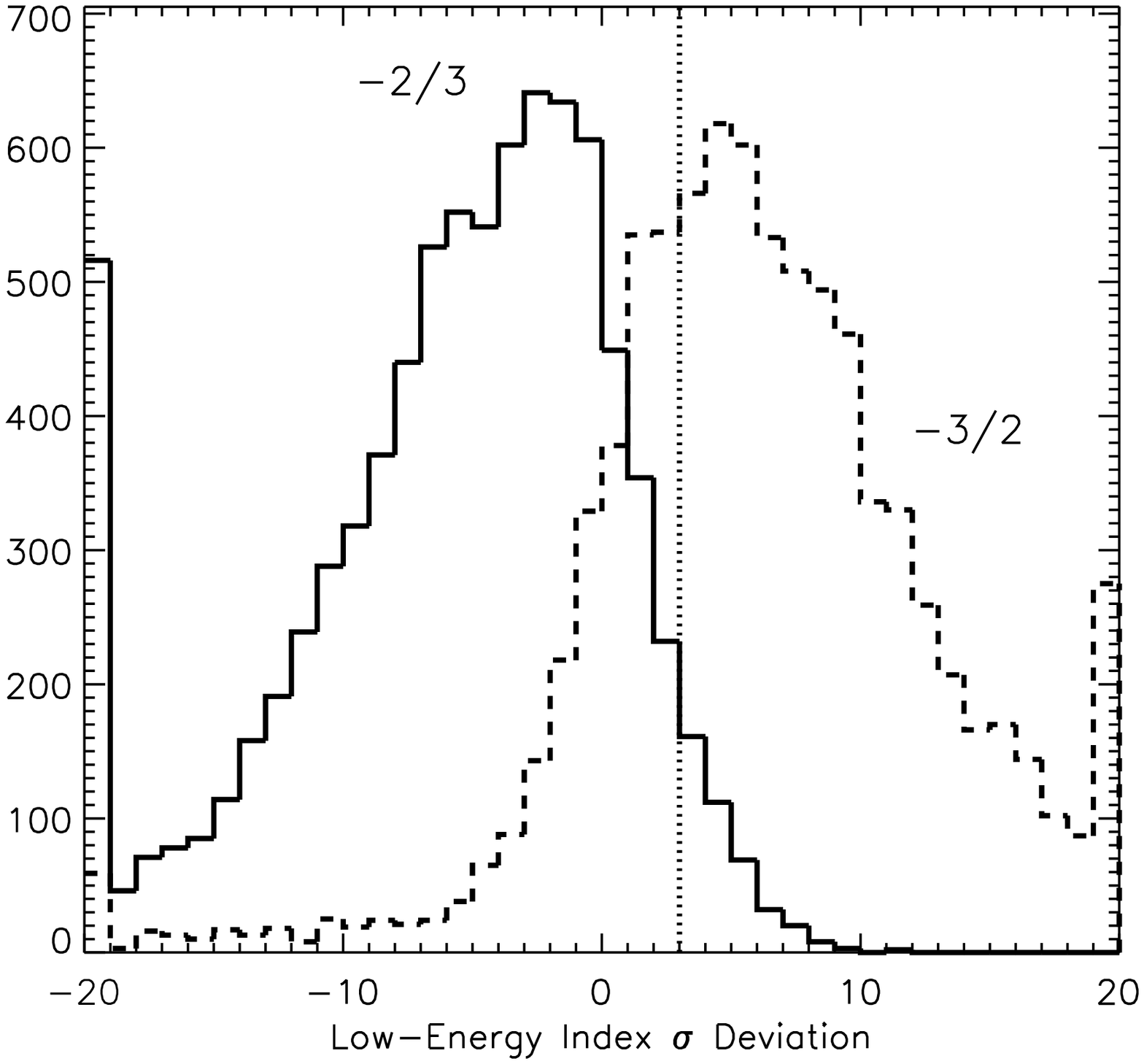}
\caption{Distribution of deviations (in units of $\sigma$) of the
low-energy indices of 8459 time-resolved spectra from values
predicted by optically-thin synchrotron emission: $\alpha = -2/3$
(solid line) and $\alpha -3/2$ (dashed line).  Positive 3$\sigma$
deviation is marked by the dotted line, i.e. spectra to the right
of the line are incompatible with synchrotron radiation. Overflow
counts at each end of the distribution are summed up in the
ultimate bins.}
 \label{fig:yuki}
\end{figure}

Indeed, a large fraction of all prompt time-resolved spectra have
significant deviation of their low-energy indices from values
predicted by optically-thin synchrotron emission. 
Figure \ref{fig:yuki} illustrates how common this is by showing
the distribution of deviations in units of $\sigma$ of 8459
time-resolved spectra from the BATSE spectral catalog of Kaneko et
al. 2006. In this analysis the $\alpha_{\rm eff}$ was used. The
fraction of fits that exceed $\alpha = -2/3$ by more than $3
\sigma$ is  4.8\% (407 out of 8459) while the fraction exceeding
$\alpha = -3/2$ by more than 3 $\sigma$ is 69.3 \% (5870 out of
8459).

\begin{figure}[!ht]
\includegraphics[width=1.0\textwidth]{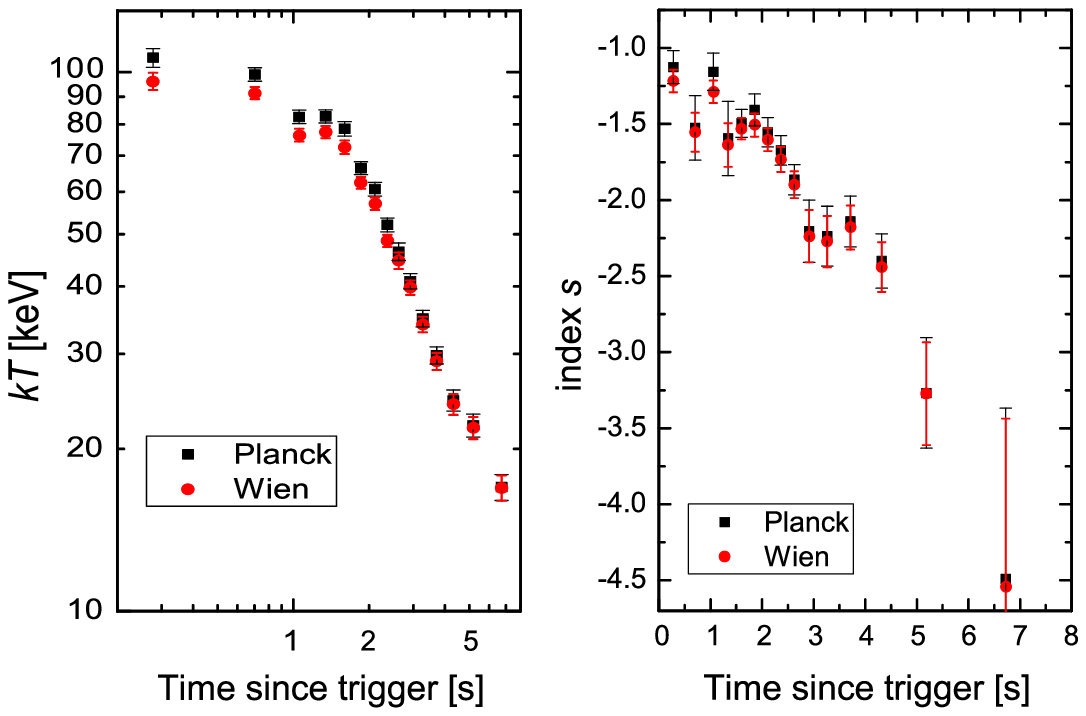}
\caption{The temperature and $s$-evolution for GRB 980306
(\#6630). The black dots (square) are for a photosphere model in
which the thermal component is described by a Planck function,
while the red dots (round) uses a Wien function. 1-$\sigma$
errors.}
 \label{fig:6630}
\end{figure}

From a theoretical point-of-view it is important to know whether
the thermal component is described by a Planck function, such as
in \citet{ryde04, ryde05} or whether it is better described by a
Wien distribution. The latter arises in a photon-starved
atmosphere, dominated by scattering (see further appendix A). The
difference between these spectra are only in the asymptotic
low-energy slope, $F_{\rm E} \propto E^2$ and $\propto E^3$,
respectively, which in general is difficult to clearly
distinguish, unless the observations have high signal-to-noise
ratio (SNR) and/or a broad energy band is used. Returning to the
spectra in \# 2193, a pure Wien function gives a total $\chi_\nu^2
(\rm{dof}) = 1.0607 (4687)$ with $q = 0.002$. The expectation
value and the variance of the $\chi^2$ distribution is 4687 and
thus its standard deviation is 68. The $\chi^2$-values are
$\chi^2_{\rm Planck} = 4678$ and $\chi^2_{\rm Wien} = 4972$, that
is, a difference of $294$. Purely statistically there is therefore
a preference for the Planck function. However, it could very well
be the case that the Wien peak is accompanied by a power-law
component. To investigate this we fit the spectral evolution with
a \citet{band93} function (with the $\alpha$-parameter free to
vary) combined with a power-law component, and find that the best
fit gives a weighted average of $<\alpha_{\rm w}> = +0.98$, but
the scatter is large; the un-weighted average is $<\alpha> =
1.54$. There are thus time bins with harder than Planck spectra.
Indeed, a fit with a Planckian plus a power-law and a fit with a
Wienian plus a power-law give similar $\chi^2$-values. The shape
of the thermal component can be compensated for by the non-thermal
component, a fact that gives rise to the ambiguity of the fits. It
is therefore not possible to clearly determine whether the thermal
component of the photosphere model is better described by a Wien
or a Planck function. However, an important point is that the
evolution of the temperature, $kT = kT(t)$, and the power-law
index, $s = s(t)$, are not greatly affected by the shape used for
the thermal component. This is also the case for bursts which have
a significant non-thermal contribution, for instance, GRB980306
(\# 6630), shown in Figure \ref{fig:6630}, where it is seen that
the $kT-$ and $s-$evolutions remain quite similar. We note that
the Planck function fits have somewhat lower $\chi^2$-values in
this case as well.

Regarding the reliability of the spectral measurements at sub-peak
energies it should be noted that it is enhanced by the fact that
the photon counts are higher at lower energies, since the spectral
photon flux typically goes as the reciprocal of the energy. This
is in contrast to the super-peak energies where the number flux
falls off rapidly. On the other hand, the instruments always have
an energy-dependent photon detection efficiency which decreases
both towards the upper and lower ends of the observed energy
window.  However, it is important to note here that, for BATSE,
the high-energy sensitivity drop is more significant compared to
the lower energies (within the analysis energy range $\geq 25 -
\sim 1800$ keV). Indeed, the effective area of the BATSE LAD at 30
keV roughly equals the one at $\sim$ 500 keV (Fishman et al.
1989).

In conclusion, the existence of {\it time-resolved} spectra that
violate the optically-thin synchrotron (isotropic pitch angles)
interpretation of prompt phase GRB spectra is therefore made
certain beyond any reasonable doubt, in particular since several
independent experiments have detected them. The BATSE cases, shown
above, clearly and significantly disfavor optically-thin
synchrotron emission as a model viable for all time-resolved
spectra. However, the data cannot clearly discriminate between
Wien and Planck shape of the thermal component, even though, there
is a slight favor for the Planckian, in that the fits mostly has a
somewhat lower $\chi^2$-value. Importantly, the temperature and
non-thermal power-law evolutions are only marginally affected. To
be able to clearly distinguish between a Wien and Planck
interpretation, data with broader energy coverage are required
since that would permit a stronger constraint to be reached of the
power-law slope of the non-thermal component.

\subsection{Significance of the decomposition}

In the photosphere model the spectra are interpreted as being
composed by two separate components. Since this model has the same
number of free parameters as the Band function the fits can easily
be compared. In most cases the fits are found to be of the same
quality (Ryde 2005). Such an example is given by GRB 911031 (\#
973), from which a spectrum is shown in the left-hand panel in
Figure \ref{fig:decomp} (the figures are plotted in $E F_{\rm E}$
units). This burst is equally well fitted by a Band function
[$\chi_\nu^2 ({\rm dof}) = 1.13(108)$] as by a photosphere model
[$\chi^2_\nu ({\rm dof}) = 1.18(108)$]. For the particular
spectrum in the figure we find that $\alpha = -1.0 \pm 0.2$,
$\beta = -1.8 \pm 0.1$ and $E_{\rm p}=360 \pm 112$, while $kT = 58
\pm 7$ keV and $s = -1.53 \pm 0.04$. However, in other cases the
decomposition is clearly necessary, for instance in the case
depicted in the right-hand panel in Figure \ref{fig:decomp} (GRB
960924; \# 5614). Here a Band model fit gives $\chi_\nu^2 ({\rm
dof}) = 8 (105)$ with $p = 0.00$ and a photosphere model fit gives
$\chi_\nu^2 ({\rm dof}) = 1.15 (105)$ with $p = 0.14$. The latter
fit gives $s= -2.2 \pm 0.1$ and $kT = 112 \pm 1$ keV. The spectrum
of GRB 960924 does not have a hard sub-peak power-law slope, as
discussed in the cases above, but rather a wavy structure in
energy, a feature that is not captured by the Band function or a
broken power-law. If this burst had been observed exclusively
above 200 keV it would probably be interpreted as being a purely
thermal burst, similar to the bursts studied in Ryde (2004).
Another example of a burst clearly needing a two-component
decomposition is given in Figure \ref{fig:decomp2} (GRB 960530;
\#~5478). Here the left-hand panel shows the results of a fit with
the Band function giving the parameters $\alpha = 1.7 \pm  1.5$
and $\beta = -2.4 \pm 0.3 $ while $E_{\rm p} = 104 \pm 15$ keV
[$\chi_\nu^2 ({\rm{dof}}) = 1.02 (109); \; p = 0.41$]. The
right-hand panel shows a fit with the photosphere model, giving
$kT = 28 \pm 2$ keV and $s = -0.62 \pm 0.27$ [$\chi_\nu^2
({\rm{dof}}) = 0.87 (109); \; p = 0.83$], with more reasonable
residuals. In this case, the Band model cannot account for the
high-energy rise in the flux beyond 600 keV. This component should
be most important at energies beyond the BATSE window studied
here. Interestingly, a few super-MeV detections have been made to
date which indicate the presence of a possible, additional
emission component at these energies \citep{hurley,atkins,gonz}.
In comparing the two plots in Figure \ref{fig:decomp2}, it is
important to note the property known as obliging, that is, for
models which fit the data badly, the data points tend to be
incorrectly depicted. This is a well-known property of the
forward-folding technique used for the deconvolution \citep{fen,
BS}. The observed quantity is the photon counts, while the
physically interesting photon flux or energy flux are derived
quantities which are model dependent. Therefore a plot showing the
photon or the energy flux can only be trusted fully if the model
fits the data well. However, the statistical fitting is made on
the count data and is independent of this effect. GRB960530 is
further discussed in the following section.

\subsection{Spectral Evolution}
\label{sec:SE}

The choice of the model used to fit the $\gamma$-ray data will
lead to distinctly different interpretations of the cause of the
apparent, observed, spectral evolution. This is illustrated in
Figure \ref{fig:829_5} which shows the spectral evolution of
GRB910927 (\# 829). The Band et al. (1993) model is used for the
fits in the upper panels whereas the photosphere model was used in
the lower-panels. There is an apparent evolution of the low-energy
part (sub-peak) of the spectrum. This gives rise to a significant
change in the measured $\alpha$-value, which varies from
approximately +1.5 to -0.5 (see Fig. 1 of \citet{crider}). Indeed
this burst is often referred to in arguing for that bursts tend to
emit in several different radiative regimes while active. This
could, for instance, be due to a variation in the opacity --
optically thick to an optically thin regime and/or due to a
variation of the pitch angle distribution of the electrons -- from
a small pitch angle to isotropic regime \citep{crider, LP00}.
However, fitting the spectra with the photosphere model, the
picture changes. The apparent spectral evolution is now simply due
to changes in temperature and index $s$, both of which behave
similarly to other bursts, that is, a very typical behavior. The
left-hand panel in Figure \ref{fig:829_2} shows the change in
temperature of the thermal component, $kT = kT(t)$, which follows
the canonical behavior found for thermal pulses, that is, a broken
power-law in time. The fit to $kT(t)$ was made with the function
derived in Ryde (2004), for which the break-scale may vary. The
fit yields that the initial power-law has an index of $-0.25 \pm
0.02$, which later breaks into a power law with index $-0.67 \pm
0.13$. The break-scale is described by the parameter $\delta = 0.4
\pm 0.2$, which is defined in equation (6) of Ryde (2004). The
time-averaged, reduced $\chi^2$ of the fits is shown in the
right-hand panel in Figure \ref{fig:829_2} and is found to be
$\chi^2_\nu (\rm{dof}) = 0.89$ (3498), which should be compared to
the Band function fits which have $\chi^2_\nu (\rm{dof}) = 0.92
(3498)$.  It is cases like this one, with an initially high
peak-energy and a hard non-thermal component, that gives rise to
pulses which have an initial thermal phase. We therefore conclude
that the strong $\alpha$-evolution, that is often attributed to
bursts, is actually an artifact of the evolving relation between
the thermal and the non-thermal components, combined with the
limited spectral energy-range of the observations.

The majority of bursts have spectra for which the  $\alpha$-index
is either a constant or becomes softer with time, like in the
cases discussed above. However, in GRB960530 (\# 5478) $\alpha$
becomes apparently harder ($\alpha$ becomes larger) which is
demonstrated in the left-hand panel in Figure \ref{fig:960530:1}.
We will now show that this power-law evolution again can be
naturally ascribed to an artifact of the empirical fit with the
Band function, and that it simply reflects the underlying
variation in the components of the photosphere model and their
relative strengths. Figure \ref{fig:960530:2} shows four
time-resolved spectra to which the photosphere model was fitted.
In the last two spectra, the power-law component is important in
determining the spectral shape at the highest energies. The
temperature decays again as a broken power-law with slopes $\sim
-0.4$ and $\sim -0.7$ (middle panel in Fig. \ref{fig:960530:1}).
The thermal flux is indeed dominant initially \citep{RB}, however,
the power-law index, $s$, behaves differently from other bursts in
that it becomes harder with time; $s$ makes a jump from $\sim
-1.5$ to $-0.67$ at approximately 5 seconds after the trigger
(right-hand panel in Fig. \ref{fig:960530:1}). The thermal
component is thereby revealed at low energies, with its hard
Rayleigh-Jeans tail. This leads to the large $\alpha$-values that
are found when the Band function is fitted to the spectra. The
change in $s$ from $\sim-1.5$ to $\sim-2/3$ can be interpreted as
that the synchrotron cooling frequency of the electrons passes
through the band-pass, towards higher energies, at around 5 s.
Such a behavior can be imagined if the amount of dissipated energy
going into the magnetic fields decreases with time.  Since the
error bars are somewhat large, this conclusions is by necessity
not beyond challenge. As can be seen in Figure \ref{fig:960530:1},
the index $s$ is also consistent of being constant at, say,
$\sim-1.5$. The whole time evolution can in fact be fitted to a
model with $s$ fixed at $-1.5$. This model then has only {\it
three} free parameters and gives an excellent fit to the whole
spectral evolution with a total $\chi^2_\nu ({\rm dof})= 0.997
(1870)$. The thermal component, on the other hand, behaves
similarly to other bursts with a broken power-law decay in time.
It should also be noted here that with a pure synchrotron model,
instead of the photosphere model, the observed behavior of
GRB960530 is difficult to explain.

\subsection{Distribution of the Power-Law Indices}

\citet{ryde05} used the photosphere model to study a sample of the
25 strongest pulses in the \citet{KRL} catalogue. In the left-hand
panel in Figure \ref{fig:histo} is shown the distribution of the
power-law indices, $s$, of the non-thermal component that was
found for all the 347 time-resolved spectra that were modelled.
Note that $s$ most often varies throughout a burst as well.
Interestingly, the distribution is peaked at $s = 1.5 - 1.6$. This
is the value expected if most spectra are due to a population of
fast cooling electrons ($s = - 1.5$) or acceleration in
relativistic shocks due to first order Fermi process; $s =
-(\hat{p} + 1)/2 = -1.6$. Here $\hat{p}$ is the power-law index of
the electron energy distribution. We also note that no cases are
harder than the slow-cooling synchrotron slope of $-2/3$. A larger
sample is presented in Battelino et al. (in prep.) with a similar
distribution. This distribution should be compared to the
right-hand panel in the figure 
which shows the distribution of the Band $\alpha$-values from the
catalogue of Kaneko et al. (2006). Here many cases have spectra
harder than $\alpha = - 2/3$ and thus the distribution crosses the
'line-of-death' of the optically-thin synchrotron model. The
distribution peaks at $\sim -0.8$, which does not have an
immediate physical meaning. Early suggestions included thermal
bremsstrahlung (with a Gaunt factor of 1) of an optically-thin hot
plasma  which yields $\alpha =-1$ \citep{RL}. However such a
mechanism is too inefficient to be able to give rise to GRB
spectra \citep{liang82}.

\subsection{Merits of the Photosphere Model}

Optically-thin synchrotron emission is excluded as a viable model
for the prompt phases in a fair fraction of the time-resolved
spectra in GRBs. Indeed a large variety of spectral shapes and
evolutions have been described in the literature. We argue that
the photosphere model gives a reasonable description, however, it
should be noted that other models have been proposed. For
instance, small pitch-angle synchrotron emission and inverse
Compton emission from mono-energetic soft photons are alternative
scenarios that give hard low-energy spectral slopes. See further
details on this in \S \ref{sec:discussion}.

We find the following main features when the photosphere model is
used to interpret the $\gamma$-ray data of GRBs.

(i)  For most cases the fits are as good as the fits made using
the Band model. In some cases the $\chi^2$-values are indeed lower
for the photosphere model.

(ii) The photosphere model can naturally incorporate all types of
spectral behaviors, even bursts which are consistent with
optically-thin synchrotron emission. In particular, it naturally
explains the existence of pulses that are thermal throughout their
duration (Ryde 2004), simply as cases for which the non-thermal
component is weak or does not affect the spectrum in the observed
energy range. Some bursts are thermal only initially, while others
are thermal only at the end. The different realizations is
naturally explained by the change in relative strength between the
two spectral components. The theoretical interpretation of the
variation in component strengths is discussed in \S
\ref{sec:T_N_T}. This means that we have one and the same model
for all bursts.

(iii) The strong $\alpha$-evolution, that is found in many bursts
when the Band model is used, has been discussed extensively in
works trying to explain the emission process at play. We argue
that this evolution in fact does not need to reflect the emission
process itself but rather the variation of the two components and
their relative strength.

(iv) The distribution of the spectral index of the non-thermal
component, here modelled by a power-law across the observed
window, is peaked at approximately -1.5. This is the value that is
expected from a population of fast cooling electrons emitting, e.g
synchrotron radiation. Furthermore, the index $\sim -1.6$ is
expected from relativistic shock accelerated electrons in the slow
cooling regime. Such a preferred value is not found from the Band
$\alpha$-distribution. The interpretation of the change in $s$ can
be that the synchrotron frequency passes through the observed
window \citep{ryde04, milanex}  or, alternatively, it can be
interpreted as the spectrum actually changes due to increased
cooling, see further discussion in Ryde et al. (2006, in prep.)

(v) The photon index of the non-thermal component are mainly
smaller than $-2/3$, that is, the 'line-of-death' for
optically-thin synchrotron emission is not crossed.

(vi) Interestingly the temperature, $kT$, evolves similarly for
most bursts, except for a few cases. This is also the case for
bursts which, a priori, do not require a thermal component, since
their spectra can equally well be fitted by a optically-thin
synchrotron model. For these burst, the temperature of the thermal
component still behave in the canonical way; having a break in the
cooling behavior.

Finally, models invoking thermal emission naturally give rise to
correlations between flux and spectral peak energy, i.e.
temperature. Such correlation are observed for an ensemble of
bursts \citep{LP00, amati} as well as during the course of one
single burst \citep{RS00, BR01}. Also the clustering of the peak
energy that is observed can be a natural consequence. These issues
are discussed in more detail in \S\S 3.3 and 3.5.

\section{Theoretical Implications of Strong Thermal Emission}
\label{sec:theory}

We interpret the thermal component as emission from the fireball
photosphere, the point at which the outflow becomes optically
thin. Quite general conclusions can then be drawn from the fact
that we detect strong thermal emission, which is shown in the next
section. In appendix B, the basic evolution of the fireball is
presented and  we will describe how it can be used to interpret
the existence of and the behavior of the thermal component.

\subsection{The Photosphere}

The three, main unknown parameters of the outflow at the point
where the thermal emission is emitted is the Lorentz factor,
$\Gamma$, the distance from the central engine, $R$, and the
comoving lepton density, $n'$. However, these properties can be
deduced using the observables, mainly the temperature, $T$, pulse
length, $t_{\rm pulse}$, total energy emitted, $E_{\rm tot}$ and
the associated solid angle, $\Omega$, into which the outflow is
collimated. The latter two quantities are related by the isotropic
equivalent energy as $E_{\rm iso} = E_{\rm tot} \times
4\pi/(2\Omega)$. In the following, we will denote the frame of
reference at rest with the progenitor as the lab frame and the
comoving frame is at rest with the outflow. Quantities in the
latter frame are primed. Finally, observed quantities are made in
the observer frame.

We have three equations to solve for the three unknowns.

(i) At a certain radius $R$ the optical depth to electron
scattering, given by
\begin{equation}
 \tau_{\rm hor} = \sigma_{\rm T} n' l'_{\rm hor},
 \label{eq:tau}
\end{equation}
reaches unity and the entrained photons expand freely towards the
observer maintaining their thermal distribution. Here,
$\sigma_{\rm T}$ is the Thompson optical depth and $l'_{\rm hor}$
is the horizon length, that is, the distance travelled by a photon
within the shell during the lab frame time $R/c$, the dynamical
time-scale. This corresponds to a comoving time $t'=R/c \Gamma$
and thus a comoving distance
\begin{equation}
 l'_{\rm hor} = \frac{R}{c \Gamma}\,c=\frac{R}{\Gamma}
 \label{eq:hor}
\end{equation}

(ii) We assume that the energy density in the black-body emission,
$U'_{\rm BB} \equiv a T'^4$  ($a = 7.56 \times 10^{-15}$ erg
cm$^{-3}$ K$^{-4}$) is equal to the radiation energy density
$U'_{\rm BB} = U'_{\rm rad}$, which in its turn is proportional to
the internal rest mass energy density $U'_{\rm rm} = n' m_{\rm p}
c^2 $, where $m_{\rm p}$ is the mass of the proton:
 $ U'_{\rm BB}
= \xi \, n' m_{\rm p} c^2 =  \xi \, m_{\rm p} c^2 \frac{\tau_{\rm
hor} \Gamma}{\sigma_{\rm T} R},
 $
where $\xi$ is the proportionality factor and includes possible
dilution factors in a scattering atmosphere. Equations
(\ref{eq:hor}, \ref{eq:tau}) was used in the last step. With
$T'=T/\Gamma$ this gives an expression for the Lorentz factor
\begin{equation}
\Gamma^5 = \frac{ T^4 a \sigma_{\rm T} }{ \xi \, m_{\rm p} c^2
\tau_{\rm hor}}\; R \label{eq:gamma5}
\end{equation}
The minimal $\Gamma$ is given by $\tau_{\rm hor} = 1$.

(iii) The total energy emitted in the shell is
\begin{equation}
E_{\rm tot} = 2 \Gamma \Delta' \Omega R^2 U'_{\rm rm}
(1+ \kappa)\\
= 2 \Gamma^3 \Delta \Omega R m_{\rm p} c^2 \frac{\tau_{\rm
hor}}{\sigma_{\rm T}} (1 + \kappa)\\
= 2 \Gamma^3 c^3 m_{\rm p} \Omega R \frac{\tau_{\rm
hor}}{\sigma_{\rm T}} t_{\rm pulse} (1+\kappa) \label{eq:Etot}
\end{equation}
where we used that the comoving shell width $\Delta' = \Gamma
\Delta$ and that the energy is divided into two jets on either
side of the burst. $\kappa$ holds the ratio between relativistic
particles/photons and cold particles. In the last step we used
that fact that $\Delta = c t^{\rm obs}_{\rm pulse}$, since the
pulse duration in this model is the light crossing time of the lab
frame width of the shell. Rewriting this we get an expression for
the distance from the central engine to the emission site as
\begin{equation}
R = \frac{E_{\rm tot} \sigma_{\rm T}}{2 \Gamma^3 c^3 m_{\rm
p}\Omega \tau_{\rm hor} t_{\rm pulse} (1+\kappa)} \label{eq:R}
\end{equation}
which, used with equation (\ref{eq:gamma5}), gives
\begin{equation}
\Gamma^8 = \frac{a\sigma_{\rm T}^2 }{4 \pi c^5 m_p^2 \;
(1+\kappa)\xi} \times \frac{T^4 E_{\rm iso}}{\tau_{\rm hor}^2
t_{\rm pulse}}.
\end{equation}
This gives that ($\tau_{\rm hor}=1$ for the photosphere)
\begin{equation}
\Gamma = 307 \left( \frac{1}{(1+\kappa)\xi } \right)^{1/8} \left(
\frac{T}{100 {\rm keV} } \right)^{1/2} \left( \frac{E_{\rm
iso}}{10^{53} {\rm erg}} \right)^{1/8} \left( \frac{t_{\rm
pulse}}{ 10 {\rm s}} \right)^{-1/8} \label{eq:17}
\end{equation}
which is a typical value one finds using the compactness argument
(see e.g. \citet{compactness}). Combining equations
(\ref{eq:gamma5}) and (\ref{eq:R}) again we also get that
\begin{equation}
R^{8/3} = \frac{\xi m_{\rm p} c^2 \tau_{\rm hor}} {T^4 a
\sigma_{\rm T}} \left[ \frac{E_{\rm tot} \sigma_{\rm T}}{2 c^3
m_{\rm p} \Omega \tau_{\rm hor} t_{\rm pulse} (1+\kappa)}
\right]^{5/3},
\end{equation}
which gives a radius of
\begin{equation}
R = 4.0 \times 10^{11} {\rm cm} \left(\frac{1}{(1+\kappa) }
\right)^{5/8} \left(\frac{E}{10^{53} } \right)^{5/8}
\left(\frac{T}{ 100 {\rm keV}} \right)^{-3/2} \left(\frac{t}{10
{\rm s} } \right)^{-5/8} \label{eq:19}
\end{equation}
This value corresponds approximately the radius of a Wolf-Rayet
star, generally assumed to be the progenitor.

It is interesting to note that equations (\ref{eq:17}) and
(\ref{eq:19}) follow directly from the existence of strong thermal
emission in the spectra. These general results now correspond to
$n' = 1.15 \times 10^{15} {\rm cm}^{-1}$, $l' = 4 \times
10^{9}{\rm cm}$, and $\Delta = 3 \times 10^{11} {\rm cm}$. Note
that $\Delta \sim R$ and $l = R \Gamma^{-2} \sim \Delta
\Gamma^{-2}$, that is, the horizon length is much smaller that the
wind width, indeed by a factor of $10^{-5}$. Furthermore, we have
that $ U'_{\rm BB} = a T'^4= 1.56 \times 10^{14} {\rm erg/cm}^{3}$
and $M_{\rm tot} =  n' m_{\rm p} dV' = 3.4 \times 10^{-7}
M_{\sun}$. The dimensionless entropy of the outflow is therefore
$\eta \equiv {E_{\rm tot}}/{M_{\rm tot} c^2} 
= 313$.

\subsection{A Toy Model}
\label{sec:toy}

A steady flow (constant energy) will give rise to constant
observer quantities, such as the temperature, since the
photosphere would lie at the same radius (see eq. \ref{eq:R23}),
and thus no spectral evolution would be present. Variation in the
outflow is therefore necessary since spectral evolution does
indeed take place. In this paper, we will focus on the
correlations between flux and peak energy (e.g. \citet{RS00, BR01,
amati, ghirlanda}) and a detailed analysis on the temporal
behavior of the temperature is deferred to a future publication.
One possibility for this variation is that the dimensionless
entropy of the material that is injected into the jet at $R_i$
(from where the outflow starts to accelerate) varies with time
\begin{equation}
\eta (t) = \frac{E_i(t)}{M_i(t)\, c^2} = \frac{U_{\rm rad,i}
(t)}{n'_i (t) \, m_{\rm p} c^2} \label{eq:etadef}
\end{equation}
Such a variation will lead to varying $R_{\rm eq} (t)$ (see eq.
[\ref{eq:Rsat}]) and thereby a varying $R_{\rm ph}(t)$ as
discussed below, reproducing the observed temperature behavior.
During the acceleration phase the comoving temperature decreases
linearly with radius, $T' \propto R^{-1}$ as discussed in Appendix
B. Therefore the radiation energy-density is given by
\begin{equation}
U_{\rm rad}(R) = a T'^4 (R) = a T_i'^4  \left( \frac{R}{R_i}
\right)^{-4} \label{eq:urad}
\end{equation}
Since $n' \propto R^{-3}$ (from mass conservation in eq.
[\ref{eq:mass}]), the rest-mass energy-density is then given by
\begin{equation}
U_{\rm rm}(R) = n'(R) m_{\rm p} c^2 =m_{\rm p} c^2 n'_i \left(
\frac{R}{R_i} \right)^{-3} \label{eq:urm}
\end{equation}
Defining $R_{\rm eq}$ as the radius at which $U_{\rm rm} = U_{\rm
rad}$, we have
\begin{equation}
R_{\rm eq} \propto  \frac{T_i'^4}{n_i'} \label{eq:25}
\end{equation}
The entropy injected into the outflow can depend differently on
the rest-mass energy and the radiation for different bursts.
Energy dissipation will increase the radiation energy-density
through various emission processes in the neighborhood of $R_i$.
If the dissipation occurs through collisions of denser regions
(e.g. shells), the amount of energy dissipation should depend
non-linearly on the volume-averaged particle density, since
several regions are involved. Furthermore, the energy extraction
processes around the newly formed black hole are, in general,
complex and the radiation processes involved are often expected to
have a strongly non-linear behavior (neutrino annihilation,
electron-positron pair cascades, etc.) which therefore could
account for such non-linear relations as well. In either scenario
such behavior can be parameterized according to the following toy
model
\begin{equation}
U_{\rm rad,i} \propto U_{\rm rm, i}^\zeta\; \Leftrightarrow \;
T_i'^4 \propto n_i'^\zeta \label{eq:toy}
\end{equation}
In other words we let $\eta(t)$ vary with time and require that
$\eta(t) = U_{\rm rad,i}(t)^{1-1/\zeta}$. Accordingly, bursts
which have large values of the power-law index $\zeta$ thus have a
flow in which the variation in entropy mainly results in a
variation in comoving temperature.

An alternative description is that $R_{\rm eq}$ varies due to
changes in the energy to mass ratio relation during the
dissipation in the outflow itself. The temperature and density
inhomogeneities can then be parameterized by $T' = T_{\rm eq, 0}'
\hat{T}'$ and $ n' = n_{\rm eq, 0}' \hat{n}'$, now assuming
constant initial flow conditions in equations (\ref{eq:urad},
\ref{eq:urm}). In this case the $R_{\rm eq}$--condition, $(T')^4
\propto n'$, gives a similar ratio as in equation  (\ref{eq:25})
with ${\hat{T}'}$ and $\hat{n}'$ instead of $T_i'$ and $n_i'$. The
following analysis and discussion will therefore be
correspondingly similar between these scenarios.

Using the toy model relation [eq. (\ref{eq:toy})] in equation
(\ref{eq:25}) thus yields
\begin{equation}
R_{\rm eq} \propto n_i'^{\zeta-1} \label{eq:req}
\end{equation}
Furthermore, since the comoving lepton density varies according to
$n'(R) = n_i \left( R/R_i \right)^{-2}$ beyond $R_{\rm eq}$,
equations (\ref{eq:tau}) and (\ref{eq:hor}) gives that
 $
\tau_{\rm hor} = \sigma_{\rm T} {n_i' R_i^2}/{R \Gamma}.
 $
The thermal photons are released from the outflow when the opacity
over a horizon length corresponds to $\tau_{\rm hor} =1$.
Therefore,
\begin{equation}
R_{\rm ph} = \sigma_{\rm T} R_i^2 \frac{n_i'}{\Gamma} \propto
\frac{n_i'}{R_{\rm eq}} \propto n_i'^{2-\zeta}\label{eq:rph}
\end{equation}
since $\Gamma \propto R_{\rm eq}$ (eq. [\ref{eq:Rsat}]). Equations
(\ref{eq:req}) and (\ref{eq:rph}) thus parameterizes $R_{\rm eq}$
and $R_{\rm ph}$ with the initial comoving lepton density $n_i'$
(alternatively $\hat{n}'$).

\subsection{The Expected Hardness-Intensity Correlation (HIC)}

During the coasting phase ($R > R_{\rm eq}$ and $\Gamma =
{\mathrm{ constant}}$) the observed peak in the spectrum, which is
directly proportional to the temperature, is given by
 $
T_{\rm ph}^{obs} = \Gamma T_{\rm ph}' = \Gamma T_{\rm eq}'\left(
\frac{R_{\rm ph}}{R_{\rm eq}} \right)^{-2/3}
 $
as discussed in Appendix B. The ratio of the photospheric radius
and the saturation radius is thus given by
\begin{equation}
\frac{R_{\rm ph}}{R_{\rm eq}} = \left( \frac{T_{\rm eq}^{obs}}
{T_{\rm ph}^{obs}} \right)^{3/2} \label{eq:ratio}
\end{equation}
which leads to that the luminosity is given by
\begin{equation}
L \propto \Sigma (T_{\rm ph} ^{obs})^4 \propto (R_{\rm
ph}/\Gamma)^2 T_{\rm ph} ^4 \propto (T^{\rm obs}_{\rm eq})^3
T_{\rm ph}  \propto (T^{\rm obs}_{\rm i})^3 T_{\rm ph}
\label{eq:lum}
\end{equation}
where $\Sigma$ is the emitting surface subtended by the opening
angle of $\Gamma^{-1}$, with $\Gamma \propto R_{\rm eq}$. We also
used the fact that during the acceleration phase
 $
T^{\rm obs}_{\rm eq} = \Gamma (R_{\rm eq}) T'_{\rm eq}= \Gamma
(R_{i}) T'_i = T_i^{\rm obs}.
 $
Furthermore, by combing equations (\ref{eq:req}) and
(\ref{eq:rph}) with equation (\ref{eq:ratio}) we find that
\begin{equation}
T_{\rm ph}^{obs} = T_i ^{\rm obs} n_i ^{(4\zeta-6)/3} \propto
T_i^{(19\zeta-24)/3\zeta}.\label{eq:TT}
\end{equation}
where we have used the toy model relation between initial
temperature and density, equation (\ref{eq:toy}). This relation
can then be inserted in equation (\ref{eq:lum}) to find
\begin{equation}
L \propto T_{\rm ph}^{\frac{28\zeta-24}{19\zeta-24}}.
\label{eq:Lexponent}
\end{equation}
Such a relation for the hardness-intensity correlation (HIC) is
indeed found for individual pulse structures. The power-law
indices are found to have a large dispersion with values centered
around approximately 2 (see Fig. 3 in \citet{BR01}). These values
are, however, for the total flux, i.e. both the thermal and the
non-thermal flux. The observed values of the exponent in equation
(\ref{eq:Lexponent}), will therefore be somewhat larger, since the
non-thermal flux increases in relative strength as the temperature
drops (see below in \S 3.5). The exponent as a function of $\zeta$
is shown by the solid curve in Figure \ref{fig:s}. The
interpretation is therefore that the majority of bursts have
$\zeta \sim 2$, while a few bursts have larger values; as the
parameter $\zeta$ tends to infinity the exponent tends to
$28/19=1.47$. For the former bursts this, for instance, would
correspond to approximately equal density regions colliding while
for the latter bursts the dissipation leads to that the variation
in radiation energy-density ($T_i$ or $\hat{T}$) is much dominant
over that in the rest-mass energy-density (or $n_i'$ or
$\hat{n}$). We also note that if $\zeta$ approaches $24/19 = 1.26$
the luminosity would change without any temperature change, that
is, without any significant spectral evolution. This is not
observed.

\subsection{Ratio between the thermal and the non-thermal
emission} \label{sec:T_N_T}

The non-thermal component of the spectra is assumed to arise at a
later time, farther out. In this optically-thin region the kinetic
energy, which is still residing in the outflow
can be dissipated by, for instance, shocks or magnetic
reconnections in which electrons are accelerated to non-thermal
distributions. These are cooled by emitting synchrotron and/or
inverse Compton emission. The non-thermal emission will therefore
be delayed compared to the thermal light curve with a time scale
that corresponds to the time it takes the shell to move from the
photospheric radius to the dissipation region.
\begin{equation}
\Delta t^{obs} = \frac{R_{\rm diss.}-R_{\rm ph}}{2c\Gamma^2} \sim
0.17 \; \mathrm{ s} \left(\frac{\Delta R}{10^{14} {\rm cm}}
\right) \left( \frac{\Gamma}{100} \right)^{-2}
\end{equation}
GRB pulses that are initially thermal are, in part, a
manifestation of such a delay. An example of such a time delay is
shown in Figure 13 in Ryde (2004). However, with a smaller $\Delta
R$, say $10^{13} \mathrm{cm}$, such a time delay would be
difficult to detect in general.

As mentioned above, Ryde (2005) showed that the variety in
spectral shapes that are seen in GRBs is, in part, attributed to a
variation in strength between the thermal and the non-thermal
component in the spectra. In the model presented here this
translates into a variation between the thermal energy and the
kinetic energy at $R_{\rm ph}$. This will depend on the relation
between the saturation radius and the radius at which the
photosphere occurs, which is given by equations (\ref{eq:req}) and
(\ref{eq:rph}) as
\begin{equation}
\frac{R_{\rm ph}}{R_{\rm eq}}  \propto T_i^{(12-8\zeta)/\zeta}
\label{eq:35}
\end{equation}
In more detail, since $kT \propto \rho^{1/3}$ we have that the
ratio between the energy density in thermal radiation and
non-thermal radiation, the latter assumed to be proportional to
the kinetic energy and thereby $\rho$, is given by
\begin{equation}
\frac{ U_{\rm BB} }{ U_{\rm non-th}} \propto \frac{ aT^4 }{ \rho }
\propto T \label{eq:scale27}
\end{equation}
as the fireball expands. If the photospheric radius is close to
the saturation radius then the thermal and non-thermal energies
are equally large (per definition) but if $R_{\rm ph} >> R_{\rm
eq}$ then the ratio will have decreased due to the scaling in
equation (\ref{eq:scale27}). This means that the relative strength
of the thermal component compared to the non-thermal component in
GRBs vary from burst to burst and within a burst due to variations
in the ratio $R_{\rm ph}/R_{\rm eq}$, i.e. the location of the
photosphere relative to the saturation radius. We therefore have
that
\begin{equation}
\frac{ L_{\rm th} }{ L_{\rm non-th} } = \frac{T_{\rm ph}}{ T_{\rm
eq} } = \left( \frac{R_{\rm ph}}{ R_{\rm eq} } \right)^{-2/3}
\propto T^{\frac{16\zeta-24}{19\zeta-24}}\label{eq:LL}
\end{equation}
where we have used equation (\ref{eq:ratio}) and in the last step
equations (\ref{eq:TT}) and (\ref{eq:35}). The dependence of the
exponent is shown by the dashed curve in Figure \ref{fig:s}. For
large values of $\zeta$ the ratio in the exponent goes as $\propto
T^{16/19}= T^{0.84}$. The dispersion of this exponent is somewhat
smaller as is evident from the figure. As the observed temperature
drops in a burst (e.g. due to varying $T_i$ or $\hat{T}$) the
non-thermal luminosity should become increasingly more important.
An example of this type of behavior, with a power-law relation
between the flux ratio and the temperature, is shown in the
right-hand panel in Figure \ref{fig:ratio} for GRB921207 (\#
2083).

\subsection{Luminosity-peak energy (Amati/Ghirlanda) relation}

In the discussion above the properties intrinsic to individual
bursts have been in focus, for instance, discussion on the
correlation between hardness and intensity (HIC; \citet{RS00,
BR01}). When studying the relationship between average properties
of a sample of bursts, normalization of the intrinsic properties
must be done. During the acceleration phase, the normalization is
given by the radius where the acceleration starts, $R_i$, so that
$\Gamma= \Gamma_i (R_{\rm eq}/R_i)$. From equation (\ref{eq:Etot})
we  have that
\begin{equation}
E_{\rm iso} = 4 \pi \Delta \Gamma^2 R^2 n' m_{\rm p} c^2 \propto
\Delta \Gamma^2 R_{i}^2 n'_{i} \propto R_{\rm eq}^2 n'_{i} \Delta
\propto \frac{T_i^8}{n'_i}\Delta \propto
T_i^{\frac{8\zeta-4}{\zeta}} \Delta
\end{equation}
where we have made use of equation (\ref{eq:25}). If we assume
that $\Delta$ does not vary significantly between bursts then we
have with equation (\ref{eq:TT}) that
 $
E_{\rm iso}  \propto T_{\rm ph}^{\frac{24\zeta-12}{19\zeta-24}}.
 $
While this is the total energy, the observed, radiated energy is a
fraction $\epsilon \propto T_{\rm ph}/T_{\rm eq}\propto
T^{(16\zeta-24)/(19\zeta-24)}$ (from eq. \ref{eq:LL}) or
 $
E_{\rm rad} = \epsilon E_{\rm iso} \propto
T^{\frac{40\zeta-36}{19\zeta-24}}
 $
which is proportional to the luminosity. For $\zeta \gta 3$ this
relation reproduces the observed luminosity--peak-energy relation
noted by \citet{amati}, however see also \citet{NP05}. Indeed, for
$\zeta$ tending to infinity the exponent is $40/19 = 2.1$. Since
the scatter around the Amati-relation is somewhat large, the
modified relation introduced by \citet{ghir_rel} can be used
instead: $E_{\rm iso} t_{\rm jet} \propto E_{\rm p}^2 \propto
T^2$, with $t_{\rm jet}$ being the jet break time, which implies
instead that the global parameter $ t_{\rm jet} \Delta$ does not
vary significantly between bursts.

In conclusion, the simplified toy model that was introduced above
reproduces the main characteristics of the observed burst
behaviors. For most cases the variation in radiation
energy-density of the outflow needs to be stronger than that in
the particle energy-density. Obviously, more work is needed to
make this description into a realistic model. However, the basic
features of the behavior are well described by the above
framework.

\section{Discussion}
\label{sec:discussion}

Even though a major part of the total energy in a GRBs is emitted
during the prompt phase,  consensus has not yet been reached
regarding the emission sites and mechanisms. For a model to
successfully describe this phase the following observational facts
has to be appropriately addressed:

\noindent (i) Existence of hard spectral slopes below the peak
energy, i.e., large values of the parameter $\alpha$, giving rise
to the line-of-death problem.

\begin{figure}[!ht]
\includegraphics[width=1.0\textwidth]{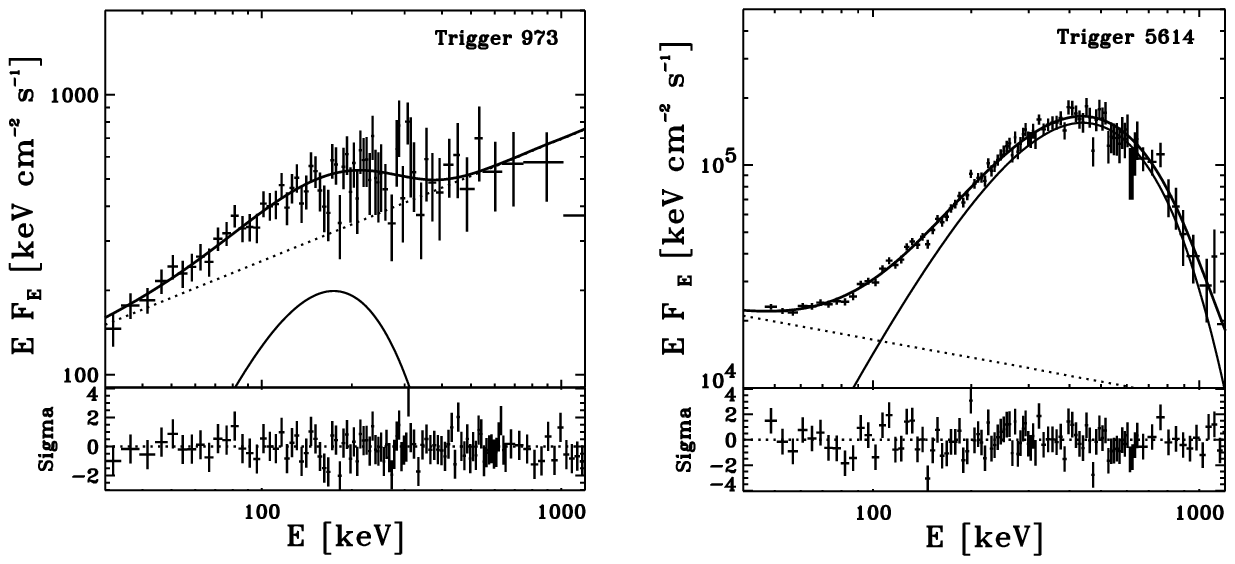}
\caption{ {\it Left panel:} A spectrum from GRB911031 (\#973; 3 s
after the trigger) fitted with the photosphere model
\citep{ryde04}, with a power-law slope of  $s = - 1.53 \pm 0.04$
and  $kT = 56 \pm 7$ keV. A fit using the \citet{band93} model
results in a similarly good fit, with $\alpha = -1.0 \pm 0.2 $ and
$\beta = -1.8 \pm 0.1$. {\it Right panel:} Time-resolved spectrum
from GRB 960924 (\#5614, $\sim 9$ s after the trigger). The burst
is totally dominated by the thermal component above 200 keV. The
two components of the photosphere model is clearly needed;
$\chi_\nu = 1.15 (105)$ compared to $\chi^2_\nu = 8 (105)$ for a
Band function fit. The spectral data points have been rebinned to
a higher signal-to-noise ratio to increase clarity. The original
energy resolution is kept in the fitting and in the residual
plots.
 }
 \label{fig:decomp}
\end{figure}

\begin{figure}[!ht]
\includegraphics[width=1.0\textwidth]{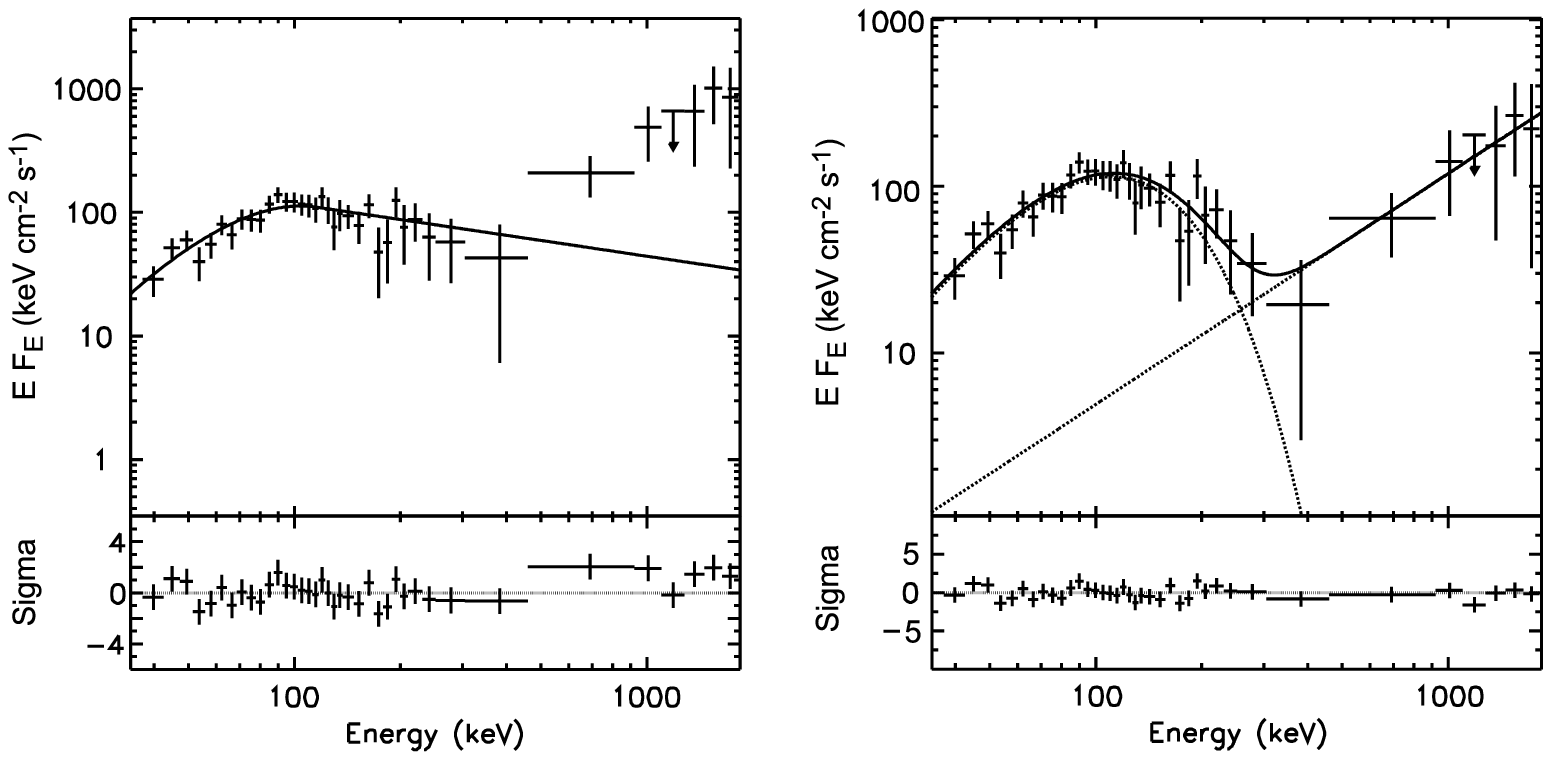}
\caption{The spectrum from GRB960530 (\#5478; 6 s after the
trigger) fitted with (left panel) the \citet{band93} model with
$\alpha = 1.7 \pm 1.5$ and $\beta = -2.4 \pm 0.3$ and (right
panel) the two-component model \citep{ryde04}, with a power-law
slope of $s = -0.62 \pm 0.27$. Note the obliging of the data
points \citep{fen, BS}.}
 \label{fig:decomp2}
\end{figure}

\begin{figure}[!ht]
\includegraphics[width=\textwidth]{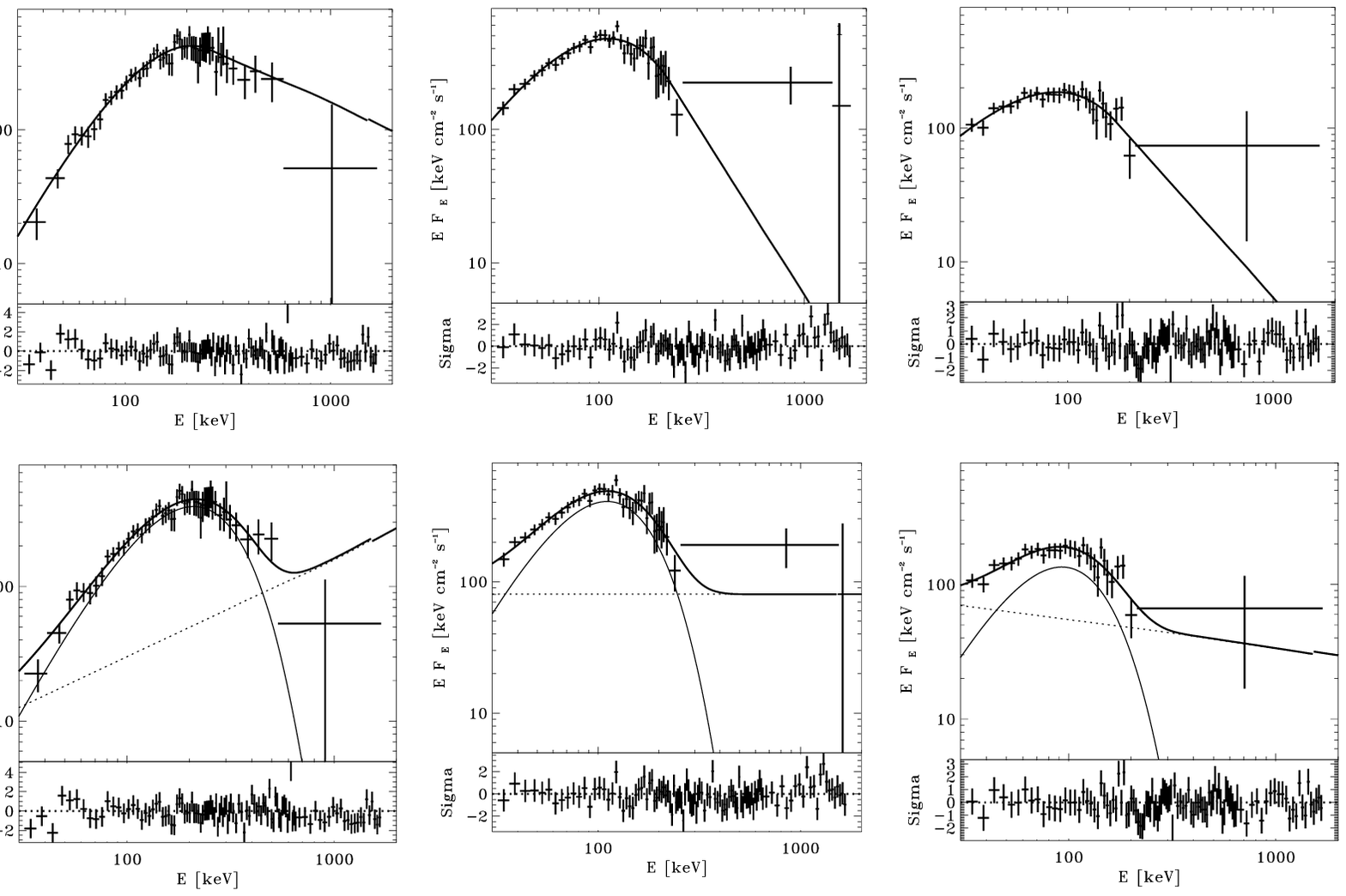}
\caption{Spectral evolution of GRB 910927 (\# 829), represented by
time bins at 1, 6, and 10 seconds after the trigger. {\it Upper
panels:} the spectral evolution that is found by using the Band
model results in evolution of $\alpha$ and $E_{\rm p}$. In
particular, the $\alpha$-evolution is noteworthy. {\it Lower
Panels:} The evolution found by using the photosphere model
becomes very typical, in particular in the evolutions of $kT$ and
$s$.  The data points have been rebinned to SNR = 3. See the text
for further details.}
 \label{fig:829_5}
\end{figure}

\begin{figure}[!ht]
\includegraphics[width=1.0\textwidth]{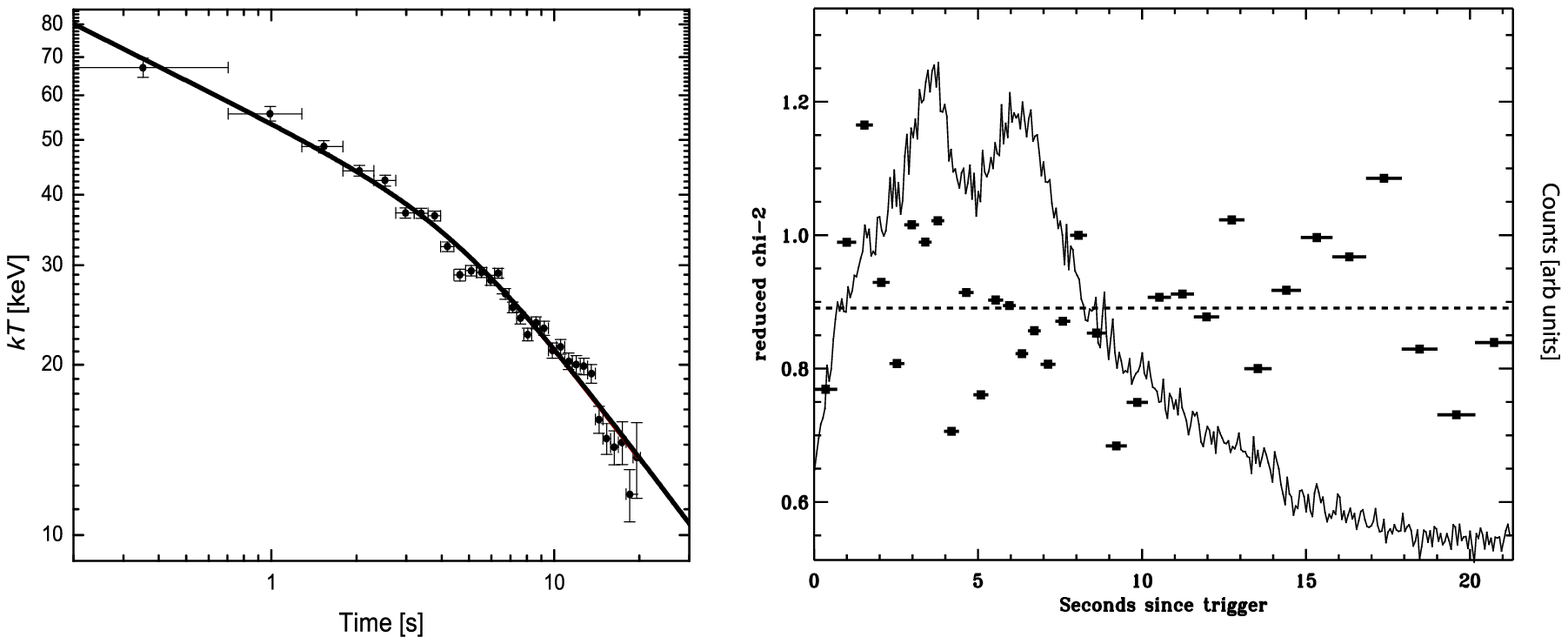}
\caption{Spectral evolution of GRB910927 (\# 829) found from using
the photosphere model. {\it Left panel:} Temperature evolution of
the thermal component. The solid line represents the best fit to a
smoothly broken power-law function. The early time power-law index
is $a = -0.25 \pm 0.02$ and the late time index $b = -0.67 \pm
0.13$. {\it Right panel:} reduced-$\chi^2$ values for the
time-resolved fits. The total value is $\chi^2_\nu = 0.89$ for
3498 degrees of freedom (dashed line). The solid line shows the
count light curve.}
 \label{fig:829_2}
\end{figure}

\begin{figure}[!ht]
\includegraphics[width=1.0\textwidth]{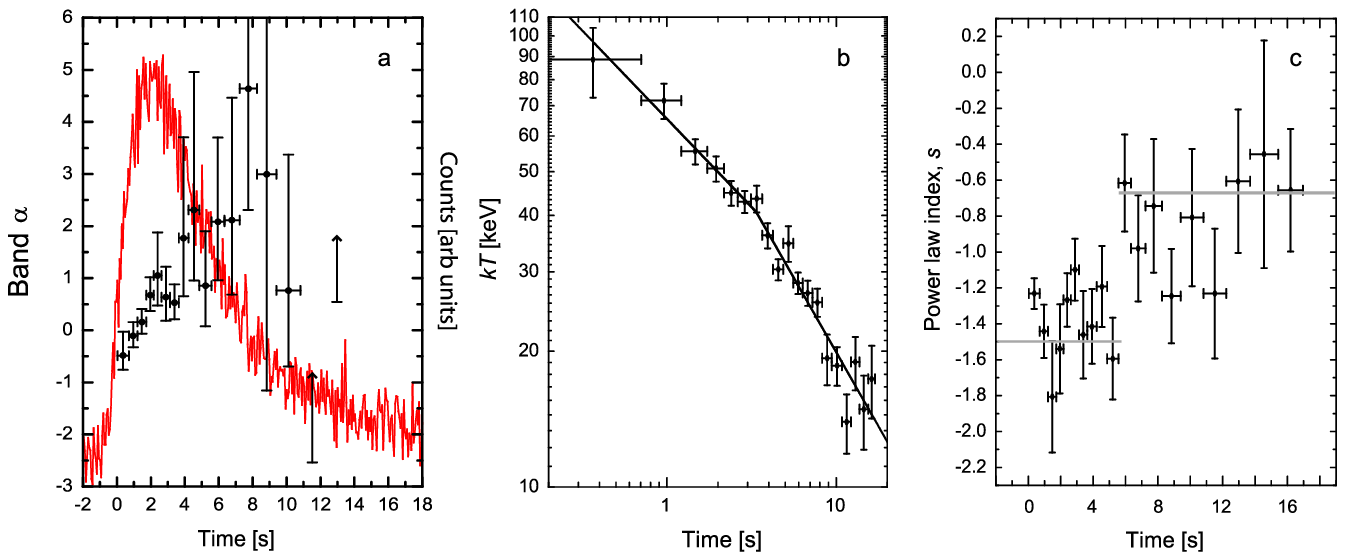}
\caption{(a) Count light curve and the low-energy power-law index,
$\alpha$. Fitting the time-resolved data of GRB960530 (\# 5478)
with the Band function, the $\alpha$ becomes harder at the end of
the burst. This is the opposite behavior to most bursts with
strong spectral evolutions. 1-$\sigma$ errors. (b) The temperature
of the thermal component evolves as a broken power-law in time.
(c) The non-thermal power-law index, $s$ makes a jump from $\sim
-1.5$ to $\sim-0.67$ at approximately 5 s. The grey lines indicate
these theoretical values.}
 \label{fig:960530:1}
\end{figure}

\noindent (ii)  Beside producing both hard and soft sub-peak
spectra, the model should be able to reproduce more complicated
spectral shapes such as the ones in Figs. \ref{fig:decomp} and
\ref{fig:decomp2}. In addition, the shape of the spectrum around
the peak that should not be too broad. The values of the model
parameter that are inferred by the observations should have a
distribution that has a reasonable explanation.

\noindent (iii) The {\it evolution} of the spectral shape. This
could be represented by parameters of non-thermal models, such as
the Band-function $\alpha$, or parameters of the photosphere
model, such as $kT$ and/or $s$. In the former case the $\alpha$
parameter translates into physical parameters depending on the
chosen  model.

\noindent (iv) The distribution of peak energies; why the peaks
mainly fall into the range they do.

\noindent (v) The correlation between peak energy and flux during
a burst, i.e. the hardness-intensity correlation, HIC, which is
characteristic for every burst \citep{RS00, BR01} .

\noindent (vi) The Amati/Ghirlanda \citep{amati, ghirlanda}
relations for an ensemble of bursts should be predictable on
reasonable grounds by the model.

\noindent (vii) Burst and pulse durations. The derived radii at
which the emission process takes place as well as the derived
Lorentz factors should not lead to the compactness problem.

\noindent (viii) The microphysics implied: does the cooling occur
in situ with the acceleration or do they occur in separated
regions? What underlying particle distributions is needed to
describe the observations, given the acceleration and dissipation
processes?

Here we have presented a physical framework that addresses most of
these points. In particular, it can explain pulses that are
thermal through out their duration. The analysis made above in \S
\ref{sec:sign} and in Ryde (2004, 2005) indicate that the
temperature has a canonical behavior with the temperature decaying
as a broken power-law, independent of the apparent behavior of the
burst. In the model presented here this simply reflects the
temperature distribution within the shell. These results are for
the class of long bursts (see e.g. \citet{balazs}) and the
corresponding behavior for short and intermediate duration bursts
\citep{Horv06} will be discussed elsewhere.

\begin{figure}[!ht]
\includegraphics[width=1.0\textwidth]{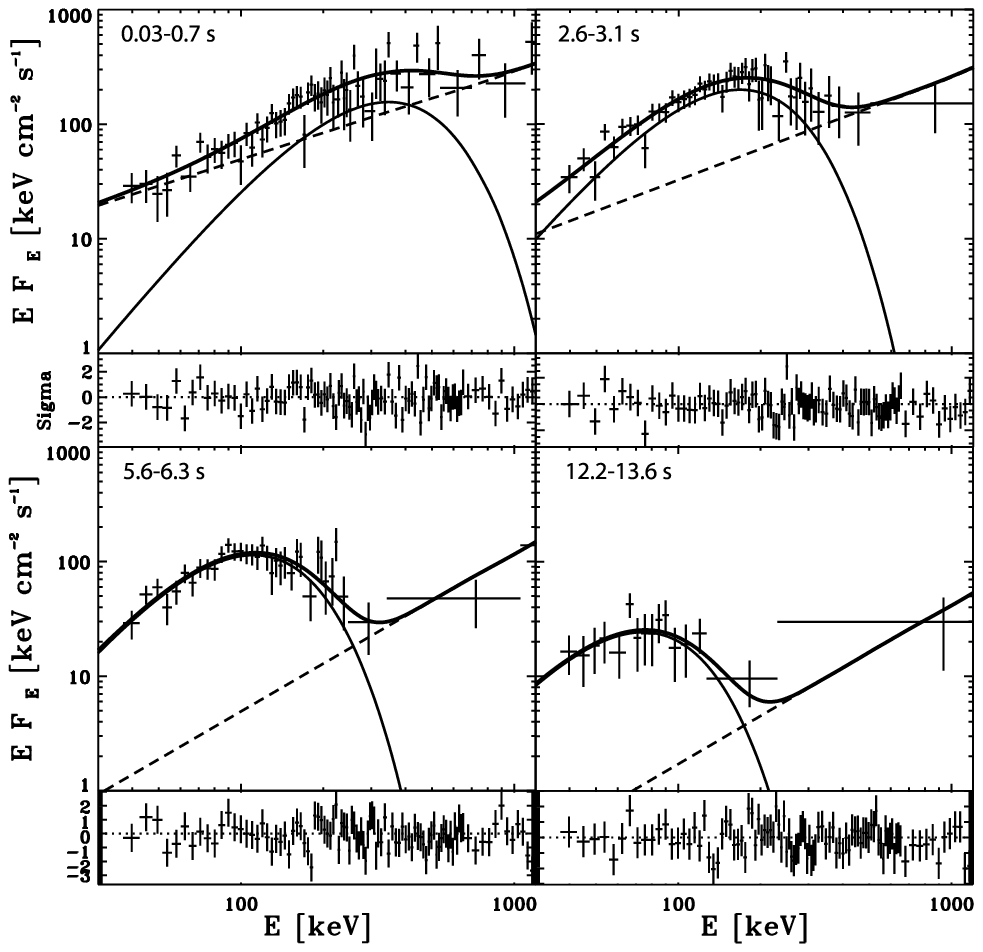}
\caption{Four time-resolved spectra of GRB960530 (\# 5478) fitted
with the photosphere model of \citet{ryde04}. The non-thermal
component is depicted by the dashed line and the time bins
(relative to the trigger) that are used for the fits are given in
the upper left-hand corner of each panel. The non-thermal spectral
component is important in determining the shape of the spectrum at
the end of the pulse.}
 \label{fig:960530:2}
\end{figure}

\begin{figure}[!ht]
\includegraphics[width=1.0\textwidth]{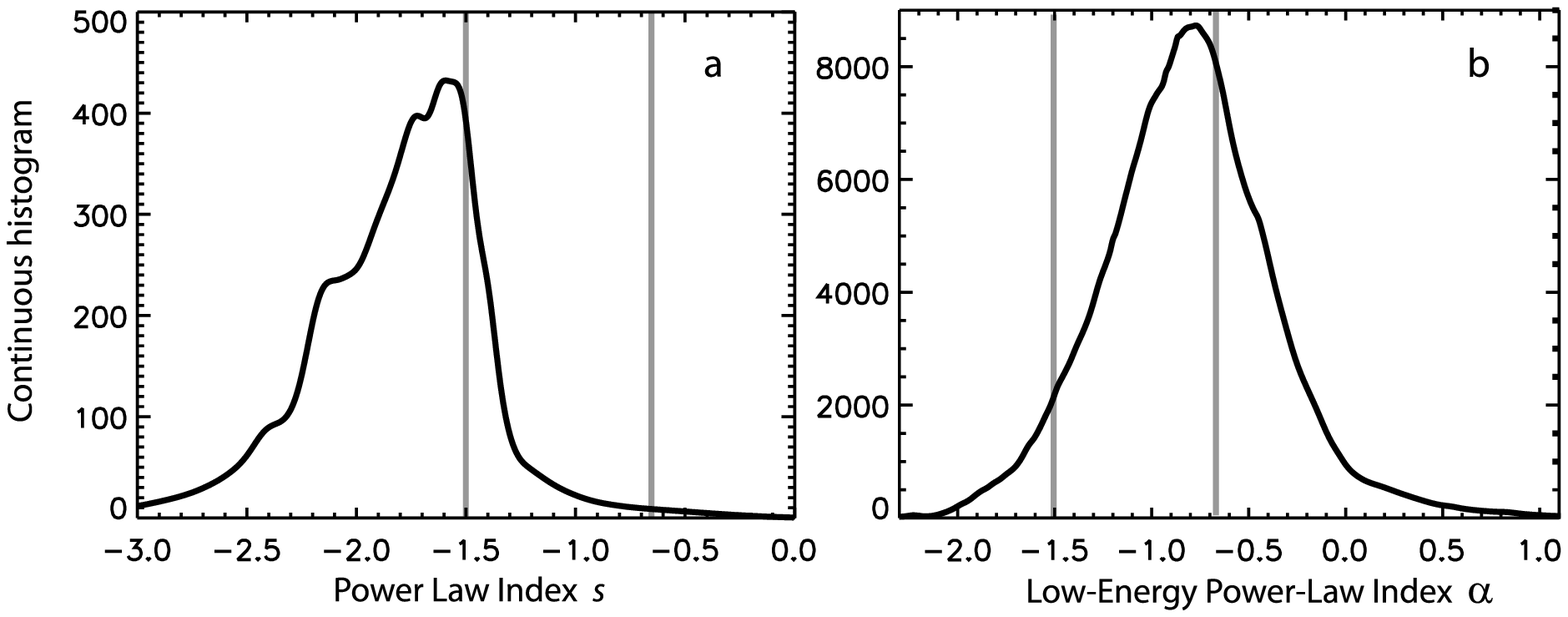}
\caption{(a) Continuous histogram of the power-law index, $s$, of
the non-thermal component of the 25 bursts studied in
\citet{ryde05}. The distribution peaks between $-1.5$ and $-1.6$.
(b) Corresponding histogram of the low-energy power-law index
$\alpha$ for the 8459 time-resolved spectra in the catalogue of
\citet{kan06}. This distribution can be compared to the ones in
Figure \ref{fig:yuki}. The grey lines indicate the indices $-2/3$
and $-1.5$ expected for synchrotron emission with slow and fast
cooling electrons. }
 \label{fig:histo}
\end{figure}

\begin{figure}[]
\includegraphics[width=1.0\textwidth]{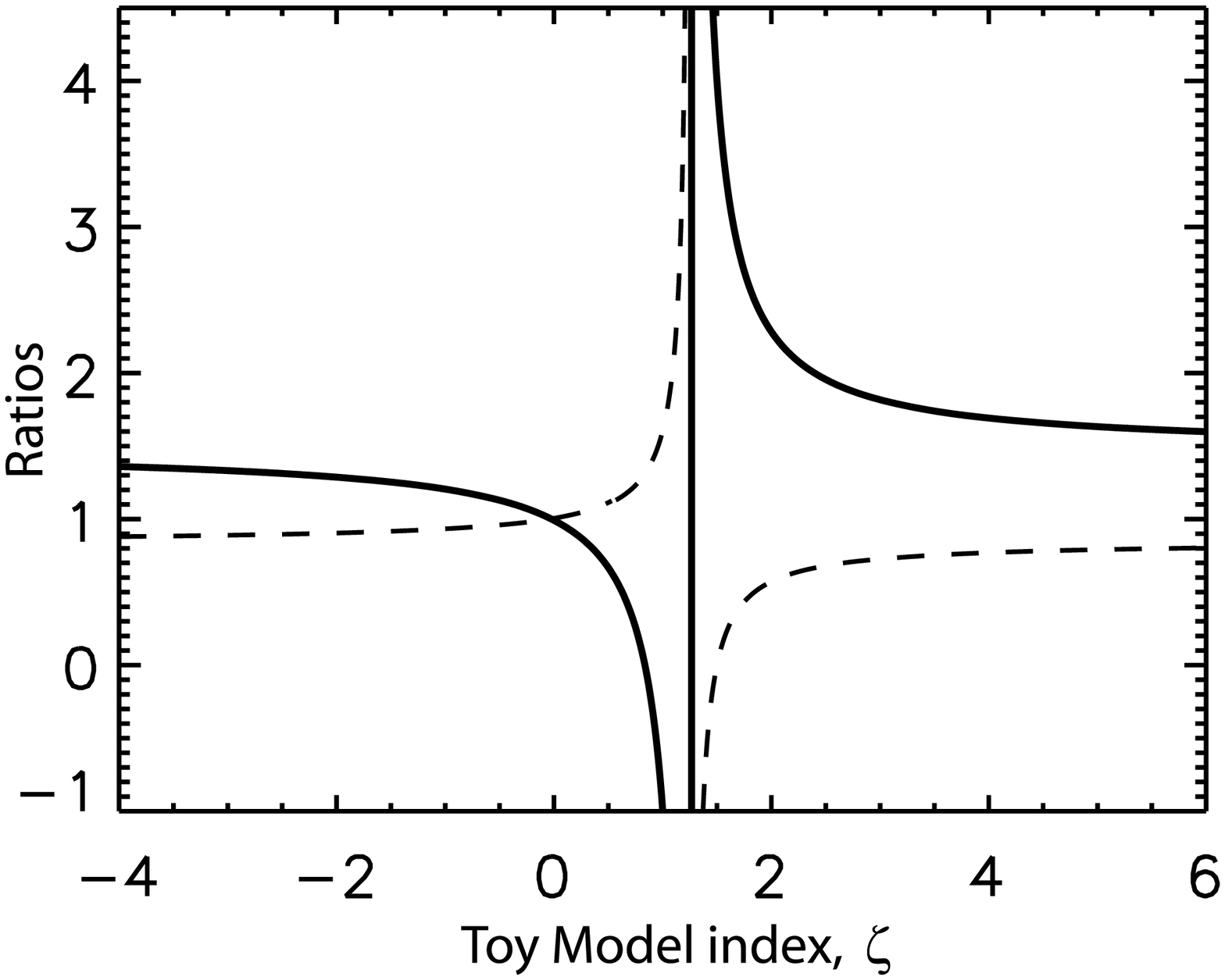}
 \caption{The behavior of the exponent in equation
 \ref{eq:Lexponent} (bold, solid line) and equation \ref{eq:LL}
 (thin, dashed line) as a function of parameter $\zeta$ of the toy model
 in equation (\ref{eq:toy}).}
 \label{fig:s}
\end{figure}

\begin{figure}[]
\includegraphics[width=1.0\textwidth]{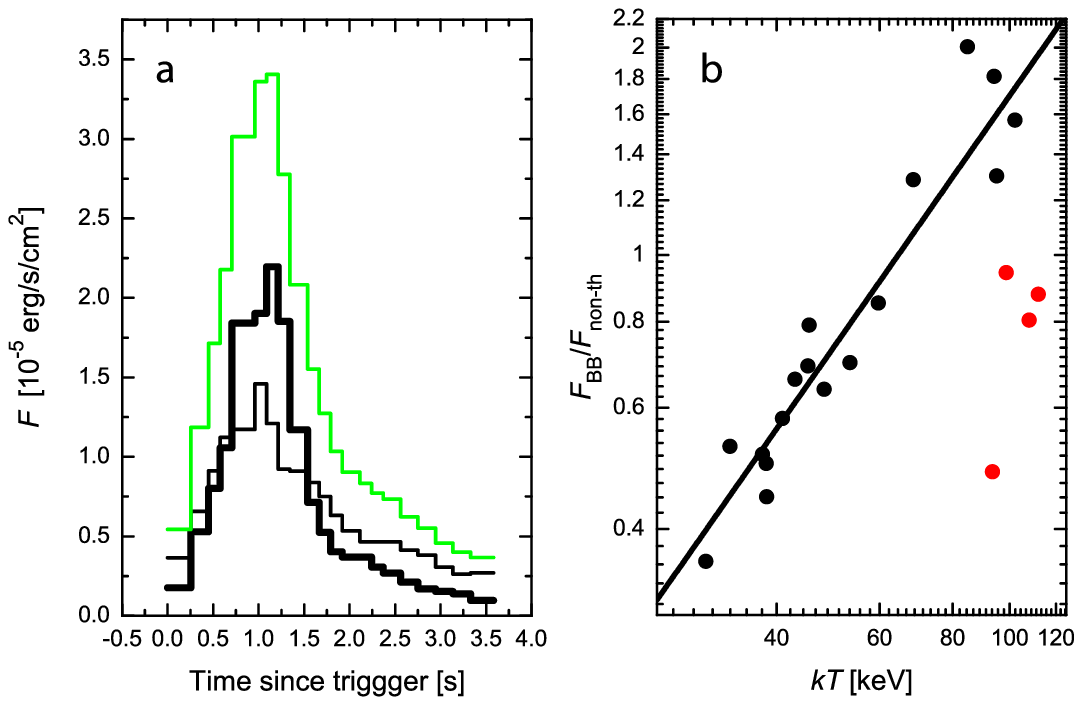}
 \caption{Example of the energy flux behavior illustrated by
 GRB921207 (\# 2083) (a) Energy fluxes in units of $10^{-5} {\mathrm{erg/s/cm^2}}$:
 Thermal (solid), non-thermal (thin), and total (grey). (b) Ratio of the thermal
 and the non-thermal fluxes as a function of temperature, $kT$. The grey points
 are for the rising phase of the pulse. The power-law relation is close to linear
 with the power-law index having a value of $1.2 \pm 0.6$.
 \label{fig:ratio}}
\end{figure}

For the bursts discussed in this paper, the peak energy was
observed to be within the BATSE energy range. However, the case
could very well be that $kT$ is actually outside of that range,
say at energies lower than 10 keV. In such a case the spectrum
within the BATSE window would mainly reflect the non-thermal
emission. The thermal component would reveal itself by an upturn
in the spectrum at low energies, similarly to the case in the
lower right-hand panel in Figure \ref{fig:960530:2}. We have also
assumed that the thermal emission is not greatly affected by a
scattering photosphere, to the extent that it pertains its
Planckian shape as indicated by the observations. However, strong
scattering can be important and has been discussed elsewhere
\citep{RM05, asaf}.

In this connection, it is worth mentioning the work by Vetere et
al. (2005) who studied rapidly varying burst which are accompanied
by a slowly varying component. They suggest that the latter
component is thermal and associated with the photosphere while the
former, variable component is due to external shocks, giving
non-thermal spectra. Such a picture fits in nicely with the model
presented above and deserves further investigation.

What alternative models does there exist? As shown above a purely
optically-thin synchrotron (OTS) emission (isotropic pitch angles)
is rejected purely by the spectral shapes. In the canonical
diffusive shock model the typical acceleration times are much
shorter than the typical synchrotron cooling times. However, the
acceleration and the cooling do not necessarily need to occur in
the same place. This will lead to electrons acquiring a fast
cooling distribution, since the cooling time is shorter that the
dynamical time scale \citep{g00}. This results in a spectral
softening, with an expected $\alpha = -1.5$ instead of $-2/3$. OTS
then becomes even less plausible. Another concern with the OTS is
that the peak energy will depend on several of the model
parameters; $\Gamma$, $\gamma_{\rm min}$, and $B_\perp$ making it
hard to get the preferred energy breaks as observed. \citet{LP00}
discussed the possibility that the hard spectral slopes are due to
synchrotron self-absorption, i.e. the electrons absorb the photons
themselves, which would lead to harder spectral slopes in the
photon fluxes; $E^1$ or $E^{3/2}$ \citep{RL}. They noted that to
get a high optical-depth to self-absorption in the BATSE energy
range somewhat extreme parameters have to be used, in particular,
one needs to invoke a very high magnetic field. At a dissipation
distance of $10^{13}$ cm, particle density of $10^8$ cm$^{-3}$,
and with  $\Gamma = 1000$ one would need approximately $10^8$ G.
Note that the generation of magnetic fields in the outflow, in
particular such strong fields, is not well understood (see e.g.
\citet{Trier}). Furthermore, the Compton $y$-parameter will in
such a case be of order unity which thus reduces the efficiency in
the observed range since emission will be up-scattered to higher
energies (Kumar 1999).

Another efficient radiative mechanism is inverse Compton
scattering: the gamma-ray spectrum could instead be an inverse
Compton image of the synchrotron spectrum, now located in the
optical since the energy increase goes as $\gamma^2$. Such a
synchrotron self-Compton model was explored by \citet{PM00}. The
low-energy slope will be the same as for the synchrotron seed
spectrum, $\alpha < -2/3$. But if the latter is self-absorbed, a
situation which now requires much lower magnetic field strengths,
the low-energy spectrum could be as hard as $\alpha = 0$
\citep{J68}.
However, as \citet{BaringB} pointed out, the spectrum around its
peak is notably broader than what is observed. Alternatively, the
inverse Compton emission could be seeded by soft photons with a
narrow energy distribution, i.e. a quasi--mono-energetic
distribution, producing a less broad spectrum and with a limiting
value again at $\alpha = 0$. Compared to synchrotron emission such
an emission mechanism also alleviates the requirements on the
magnetic field strength.

Yet another alternative is the small pitch-angle
synchrotron-emission (SPA) \citep{E73, EP73} and similarly jitter
radiation \citep{Med00, Med06}, which produce hard spectral slopes
as well, $\alpha < 0$. Here the $\gamma$-dependant distribution of
pitch angles is needed as well as a scenario to set up a
distribution for which the pitch angles are not much greater than
$\gamma^{-1}$. This could be done by rapid cooling and inefficient
diffusion transverse to the magnetic field \citep{LP00}.

The latter two models are thus the main contenders for describing
the prompt phase with a purely non-thermal model. Diffusive shock
acceleration (Fermi processes) is assumed to accelerate electrons
from a shocked thermal population into a non-thermal distribution,
which cool by these mechanisms and emit the observed radiation.
\citet{BaringB} noted, however, that both these emission processes
need an almost purely non-thermal electron distribution to be able
to fit the observed spectra. This is difficult to reconcile with
shock models, which often have a strong contribution of a thermal
population. We note here that the non-thermal component in the
photosphere model, presented in this paper, does not have its
break energy within the BATSE window. Therefore such fitting
constraints are not found and a thermal component of the electron
population can very well be present in that case.

It will be necessary to have a broader energy coverage to be able
to reach firm conclusions on the actual spectral shape in the X-
and  $\gamma$-rays, and thereby conclusively identifying the
emission process. The {\it GLAST} satellite is expected to provide
better clues with its much broader energy range compared to
presently available observations.

\begin{acknowledgements}
We wish to thank Drs. Asaf Pe'er, Stefan Larsson, and Attila
M\'esz\'aros for useful discussions and comments on the
manuscript. Financial support for this research was given by the
Swedish Research Council, the Swedish National Space Board,
Swedish Foundation for International Cooperation in Research and
Higher Education (STINT), NASA through contract NAG 5-13286 and
NSF under grant AST 03-07376.
\end{acknowledgements}

\begin{center}
 APPENDIX A
 \vskip 2mm
 SPECTRUM OF THE THERMAL COMPONENT
\end{center}
 \label{sec:th-spectrum}

In the photosphere model, discussed above, the thermal component
is the dominating feature of the spectrum, dictating the peak
evolution. Its energy flux spectrum has a general form described
by
\begin{equation}
F_E(T) = \frac{2}{h^2c^2}\,\,\frac{E^3}{e^{E/kT-\mu}-1},
\label{eq:therm}
\end{equation}
where $T$ is an effective temperature, $k$ is Boltzmann's constant
$= 1.38 \times 10^{-16}$ erg K$^{-1}$, and $\mu$ is the
dimensionless chemical potential. For photon statistics with
unrestricted amount of photons $\mu=0$ and the photon distribution
will reach a thermodynamic equilibrium so that equation
(\ref{eq:therm}) reduces to the Planck distribution, $F_E \propto
E^3 [{\exp(E/kT)-1}]^{-1}$ \citep{Planck}, which for $E<<kT$ has
an asymptotic behavior of $\propto E^2$. However, in a pure
scattering atmosphere, for instance, the photons can gain energy
from hot electrons via inverse-Compton scattering. The spectrum
will then be distorted since the number of photons has to be
conserved, that is, photons are neither created nor destroyed in
the scattering process. This restriction (Bose-Einstein
statistics) leads to a non-zero (negative) chemical potential
$\mu$ and the photons will arrive at a statistical equilibrium
instead. For $E > kT (1 + \mu)$, we can neglect the unity in the
denominator and it simplifies to the Wien distribution $F_E
\propto E^3 [{exp(-E/kT)}]$ \citep{Wien}, which asymptotically
approaches $\propto E^3$ towards lower energies. This is thus the
resulting spectrum that is established if there is a mismatch
between the number of photons and the energy that is distributed
among them.

\begin{center}
 APPENDIX B
 \vskip 2mm
 FIREBALL EVOLUTION
\end{center}
 \label{sec:fireball}

We envision the following scenario in which GRBs are formed. As
the black hole is created during the core collapse of the
progenitor star (most probably a Wolf-Rayet star\footnote{expected
to be the progenitor star of the special type of Supernova Ic that
is referred to as a collapsar}), gravitational and rotational
energy is extracted and creates a hot, optically thick gas. The
large amounts of energy per baryon involved will make it quickly
expand under its own pressure, accelerating the baryons to
relativistic velocities (Paczy\'nski 1998). This requires that the
surrounding material is not too thick for the fireball to be
halted. The collapsar model provides a depleted funnel through the
progenitor, which is created along the rotation axis since angular
momentum in the equatorial region prevents the matter to collapse
on a short time scale. In addition a precursor outburst can clear
the way for subsequent outflows. The funnel will focus the
fireball into a collimated outflow with an opening angle of
$\Omega$.

The energy injection at the central source, or engine, lasts over
a period of $\Delta/c$ where $\Delta$ is the lab frame width of
the shell that is thus emitted. The rate of injection can vary on
a shorter time scale than this and is parameterized by variations
in the dimensionless entropy $\eta$ (defined in eq.
[\ref{eq:etadef}]) or equivalently in the temperature, $T'$ and
the density of baryons, $n'$. Conservation of mass is described by
\begin{equation}
\Sigma \rho' \Gamma v = \dot{M} \label{eq:mass}
\end{equation}
where $\Sigma$ is the emitting surface, $\dot{M}$ is the rate of
baryonic matter that is injected to the outflow, $\rho'$ is the
rest mass density, and $v$ is the velocity. Conservation of energy
gives
\begin{equation}
 \Sigma (U' + P') \Gamma^2 c = H
 \label{eq:energy}
\end{equation}
where $P'$ is the pressure, and $H$ is the rate of thermal energy
ejected by the central engine (erg/s) which, during the course of
the flow, is transformed into kinetic energy. 
If we assume the outflow to be a relativistic ideal
fluid, for which the internal energy density is $U' = 3 P'> \rho'
c^2$, we can simplify equation (\ref{eq:energy}) to
\begin{equation}
 U'= \frac{3}{4}\, \frac{H}{\Sigma \Gamma^2 c}
 \label{eq:u2}
\end{equation}
which together with the conservation of mass (eq. \ref{eq:mass})
gives
\begin{equation}
 \frac{U'}{\rho'} = \frac{3}{4}\, \frac{H}{\dot{M}
 \Gamma}.
 \label{eq:u3}
\end{equation}
For a photon gas (with $\tau >>1$) we have that $U'\propto
\rho'^{4/3}$ so equation (\ref{eq:u3}) gives that $U'^{1/4}
\propto \Gamma^{-1}$. Since $U' = a T'^4$, the temperature, in the
observer frame would be $T^{\rm obs} = T' \Gamma = {\rm
constant}$. If this relation is used together with equation
(\ref{eq:u2}), which is equivalent to $T' \propto (\Gamma^2
\Sigma)^{-1/4}$ we find that $T'^{} \propto \Sigma ^{-1/2} \propto
R^{-1}$ and $\Gamma \propto R^1$. During the acceleration period
the Lorentz factor increases linearly with radius. The typical
temperature during the acceleration phase 
would be (independent of $R$)
\begin{equation}
kT^{\rm obs}\sim \frac{k}{1+z} \left( \frac{H}{4 \pi  R_i^2 c a}
\right)^{1/4} =100 \,\,{\rm keV} \left(\frac{1+z}{2} \right)^{-1}
 \left(\frac{H}{10^{51}{\rm{erg/s}}} \right)^{1/4}
 \left(\frac{R_i}{10^{8}{\rm{cm}}} \right)^{-1/2},
\end{equation}
assuming an initial radius where the acceleration starts where
$\Gamma=1$, $R_i$. As above we define the distance $R_{\rm eq}$ as
the radius at which the rest mass energy density and radiation
energy density are equal (assuming $\tau >> 1$). The flow
saturates not far beyond this radius at $R_{\rm sat} \sim R_{\rm
eq}$ and coasts along with a constant Lorentz factor. We then have
that
\begin{equation}
\Gamma = \eta = \Gamma_i \frac{R_{\rm sat}}{R_i} \sim \frac{R_{\rm
eq}}{R_i} \label{eq:Rsat}
\end{equation}
with $\Gamma_i \sim 1$. During the coasting phase we also have
that
 $
U' = \rho' c^2 + 3 P' \sim \rho' c^2
 $
which gives that $\Gamma \sim {\rm constant}$ since equation
(\ref{eq:u3})  gives that $U' \propto \rho' / \Gamma$. Mass
conservations also gives that $\rho' \propto R^{-2} \Gamma^{-1}$
which combined with $kT' \propto \rho'^{1/3}$ (the adiabatic
relation for a photon gas) gives
\begin{equation}
kT^{\rm obs} = k T' \Gamma \propto R^{-2/3} \label{eq:R23}
\end{equation}
since $\Gamma$ is constant.

\clearpage

\end{document}